\documentclass[preprint]{aastex}
\slugcomment{To appear in {\it The Astrophysical Journal}} 

\shorttitle{Asphericity in SN 1998S}
\shortauthors{Leonard et al.}

\shorttitle{Asphericity in SN 1998S}
\shortauthors{Leonard et al.}

\begin{document}

\title{Evidence for Asphericity in the Type IIn Supernova 1998S}

\vspace{2cm}

\author{Douglas C. Leonard, Alexei V. Filippenko, Aaron
J. Barth\altaffilmark{1}, and Thomas Matheson}

\affil{Department of Astronomy, University of California, Berkeley,
California 94720-3411}\affil{Electronic mail: (dleonard, alex,
tmatheson)@astro.berkeley.edu, abarth@cfa.harvard.edu}

\vspace{1cm}

\altaffiltext{1}{Present address: Harvard-Smithsonian Center for Astrophysics,
60 Garden Street, Cambridge, MA 02138.}  

\begin{abstract}

We present optical spectropolarimetry obtained at the Keck-II 10-m telescope on
1998 March 7 UT along with total flux spectra spanning the first 494 days after
discovery (1998 March 2 UT) of the peculiar type IIn supernova (SN) 1998S.  The
SN is found to exhibit a high degree of linear polarization, implying
significant asphericity for its continuum-scattering environment.  Prior to
removal of the interstellar polarization, the polarization spectrum is
characterized by a flat continuum (at $p \approx 2\%$) with distinct changes in
polarization associated with both the broad (symmetric, half width near zero
intensity $\gtrsim 10,000$ km s$^{-1}$) and narrow (unresolved, full width at
half maximum $< 300$ km s$^{-1}$) line emission seen in the total flux
spectrum.  When analyzed in terms of a polarized continuum with unpolarized
broad-line recombination emission, an intrinsic continuum polarization
of $p \approx 3\%$ results, suggesting a global asphericity of $\gtrsim 45\%$
from the oblate, electron-scattering dominated models of H\"{o}flich (1991).
The smooth, blue continuum evident at early times is shown to be inconsistent
with a reddened, single-temperature blackbody, instead having a color
temperature that increases with decreasing wavelength.  Broad emission-line
profiles with distinct blue and red peaks are seen in the total flux spectra at
later times, suggesting a disk-like or ring-like morphology for the dense ($n_e
\approx 10^7 {\rm\ cm^{-3}}$) circumstellar medium, generically similar to what
is seen directly in SN 1987A, although much denser and closer to the progenitor
in SN 1998S.  Implications of the circumstellar scattering environment for the
spectropolarimetry are discussed, as are the effects of uncertainty in the
removal of interstellar polarization; the importance of obtaining multiple
spectropolarimetric epochs to help constrain the interstellar polarization
value is particularly stressed.  Using information derived from the
spectropolarimetry and the total flux spectra, an evolutionary scenario for SN
1998S and its progenitor are presented.

\end{abstract}

\medskip
\keywords {circumstellar matter --- polarization --- stars: mass-loss ---
supernovae: individual (SN 1998S) --- techniques: polarimetric}

\section{INTRODUCTION}

Since extragalactic supernovae (SNe) are unresolvable during the very early
phases of their evolution, explosion geometry has been a difficult question to
approach observationally.  Traditionally assumed to be spherically symmetric
(for recent asymmetric models, see Burrows, Hayes, \& Fryxell 1995 and
H\"{o}flich, Wheeler, \& Wang 1999), several pieces of indirect evidence have
begun to cast doubt on this fundamental assumption of supernova (SN) theory.
Long after the explosion, evidence for asymmetry abounds, coming most
persuasively from the Galactic distribution (Lyne 1998) and high velocities (up
to 1600 km s$^{-1}$; c.f., Cordes \& Chernoff 1998) of pulsars, the compact
remnants of massive stars that have exploded.  As first pointed out by Shapiro
\& Sutherland (1982), polarimetry of a young SN is a powerful tool for probing
its geometry.  The idea is simple: A hot young SN atmosphere is dominated by
electron scattering, which by its nature is highly polarizing.  Indeed, if we
could resolve such an atmosphere, we would measure changes in both the position
angle (PA) and strength of the polarization as a function of position in the
atmosphere (e.g., Rybicki \& Lightman 1979, p. 104).  For a spherical source
which is unresolved, however, the directional components of the electric
vectors cancel exactly, yielding zero net linear polarization.  If the source
is aspherical, incomplete cancellation occurs; the resulting degree of net
linear polarization varies with the amount of asphericity, as well as with the
viewing angle and the extension and density of the electron-scattering
atmosphere.

The most recent, general study of the polarization expected from an aspherical
SN atmosphere is that of H\"{o}flich (1991), in which the continuum
polarization resulting from an axially symmetric (either oblate or prolate),
electron-scattering dominated photosphere is calculated as a function of the
photosphere's geometry and size relative to the envelope, the scattering
atmosphere's optical depth and density profile, and the axial ratio of the
envelope.  Of particular interest is the lower bound placed on the asphericity
implied by polarization measurements: By fixing all other parameters to
maximize the resulting polarization (i.e., scattering optical depth = 1,
continuum region physically small compared with envelope, object viewed
equator-on), the maximum polarization as a function of axis ratio is produced
(H\"{o}flich 1991, his Fig. 4), with polarizations ranging from 0\% (spherical)
to over 4\% (axial ratio $< 0.4$).  Determination of intrinsic SN continuum
polarization thus leads to a lower limit on the implied asphericity of the
scattering atmosphere from this model.

Previous SN polarization studies have indicated significant intrinsic
polarization ($p\approx1\%$) for core-collapse events, and small or zero
polarization for SNe Ia (Wang et al. 1996; Wang, Wheeler, \& H\"{o}flich 1997).
However, due to the difficulty of obtaining the requisite signal with even 4-m
class telescopes, most of the published results do not resolve specific line
features, which potentially contain a wealth of information about the geometry
and nature of the scattering environment.  [Notable exceptions are SN 1993J
(Trammell, Hines, \& Wheeler 1993; Tran et al. 1997) and SN 1987A (see Jeffery
1991a and references therein).]  With the larger aperture telescopes now (or
soon to be) available (e.g., Keck, VLT), prospects for improving the situation
are good.  We thus began a program to obtain spectropolarimetry of nearby SNe
at the Keck Observatory, and here report results from a spectropolarimetric
observation of the unusual core-collapse object, SN 1998S.

Classified as type IIn, SN 1998S joins the growing number of SNe whose
distinguishing observational characteristic is strong, often
multi-component emission lines lacking the broad P-Cygni absorption
features typical of ``normal'' SNe II (Schlegel 1990; Filippenko 1991;
see Filippenko 1997 for a review).  While the simple model of a red
supergiant exploding into a near-vacuum [inferred progenitor mass-loss
$\dot{M} < 10^{-5} {\rm\ M_\odot\ } {\rm yr}^{-1}$ (Chugai 1994)]
sufficiently explains the early-time spectrum of most core-collapse
SNe, spectra of young IIn events suggest intense interaction between
the ejecta and a dense circumstellar medium (CSM).  Although details
vary among objects, a physical picture is emerging in which the
progenitor star undergoes significant mass-loss, perhaps occurring in
several episodes with $\dot{M} $ as high as $ 10^{-3} {\rm\ M_\odot\ }
{\rm yr}^{-1}$ (Chugai 1997a; Benetti et al. 1999), sometimes lasting
right up to the time of explosion.  The expanding ejecta then interact
with this CSM, producing the line radiation which dominates the
optical spectrum (see Chugai 1997b for a review).

In addition to the geometric information revealed by spectropolarimetry, total
flux spectra can lend further insight into the nature of the progenitor star's
CSM, which in turn reveals the mass-loss history of massive stars in their last
stages of evolution: Do they blow off a spherical wind or one characterized by
an asymmetric morphology?  Recent, detailed images of red supergiant stars
(Monnier et al. 1999) and planetary nebulae (Sahai \& Trauger 1998) suggest
that asymmetric (though possibly axisymmetric) mass-loss in the late stages of
stellar evolution is quite common.  In fact, it is difficult to find {\it any}
observational evidence supporting spherically symmetric mass-loss; instead, a
decidedly non-spherical, often bipolar morphology is deduced (see, e.g., Soker
1999a).  If such late-time asymmetric mass-loss occurs in even more evolved
massive stars, then it follows that the immediate CSM of some core-collapse SNe
may be distinctly non-spherical, a situation whose spectral and
spectropolarimetric consequences are investigated in this paper.

Located in the Sc galaxy NGC 3877 (Fig. 1), SN 1998S was discovered on 1998
March 2.68 UT by Z. Wan during the Beijing Astronomical Observatory Supernova
Survey (Li et al. 1998; Qiu et al. 1998b; note that UT dates are used
throughout this paper).  At the time, it had an unfiltered magnitude $m=15.2$
mag, but on about March 20 it reached a peak visual magnitude of $m_V\lesssim
12$ mag and $M_B \approx 12.3$ mag (c.f., Granslo et al. 1998; Garnavich et
al. 1998c).  This made it the second brightest SN of the year, eclipsed only by
SN 1998bu (c.f., Suntzeff et al. 1999; Jha et al. 1999). At an estimated
distance of 17 Mpc (Tully 1988), SN 1998S thus reached $M_B \approx -18.8$ mag
(uncorrected for extinction), making it one of the most intrinsically luminous
type II events (c.f., Patat et al. 1993). Since it was discovered roughly 18
days before maximum brightness and a pre-discovery frame exists on 1998
February 23.7 in which no SN is apparent down to a limiting unfiltered
magnitude of $m\approx18$ mag (Li, private communication 1999), it is likely
that SN 1998S was discovered within a few days of shock breakout.

As spectra were taken, SN 1998S quickly established itself as a peculiar
variant of the already heterogeneous SN IIn subclass: In addition to smooth,
symmetric lines of hydrogen and helium superposed on a blue continuum
(Filippenko \& Moran 1998), strong \ion{N}{3}, \ion{C}{2}, \ion{C}{3}, and
\ion{C}{4} emission was seen (Garnavich et al. 1998b), reminiscent of the
``Wolf-Rayet'' features also observed in SN 1983K early in its evolution (c.f.,
Niemela, Ruiz, \& Phillips 1985).  Interestingly, SN 1983K was similarly
luminous, reaching $M_B \approx -19.5$ mag at maximum (Patat et al. 1993).  Our
spectropolarimetric observation of SN 1998S, taken just five days after
discovery, displayed a smooth, well-defined continuum with distinct line
features, making interpretation more straightforward than for objects observed
at later times.

This paper presents the early-time spectropolarimetric observation of SN 1998S
along with total flux spectra sampling the first 494 days after discovery.  We
discuss the observations and reduction techniques in \S 2 and estimate the
reddening to SN 1998S in \S 3.1.  Its unusual spectral evolution is then
presented in a series of total flux spectra (\S 3.2), with particular attention
given to its early development.  By comparing the early-time continuum shape
with reddened blackbodies, we find an excess flux beyond a blackbody continuum
at blue wavelengths.  The spectropolarimetric observation is analyzed in \S
3.3, where we attempt to remove the interstellar polarization (ISP) from the
total polarization data using three methods: First, an upper limit for the ISP
is set by reddening considerations; next, an ISP value is derived by assuming
the model of a polarized continuum with unpolarized broad lines; finally, a
different ISP is derived by adopting a similar model but with assumed
unpolarized narrow lines.  The relative merits of the different models are then
discussed in \S 4, along with a reexamination of the blue flux excess viewed in
light of our preferred polarization model.  By combining information derived
from the spectropolarimetry and total flux spectra, we conclude by presenting a
possible evolutionary scenario for the 1998S system.

\section{OBSERVATIONS AND REDUCTIONS} 

SN 1998S was observed with the Low Resolution Imaging Spectrometer (LRIS; Oke
et al. 1995) in polarimetry mode (LRIS-P; Cohen 1996) at the Cassegrain focus
of the W. M. Keck-II 10-m telescope on 1998 March 7.  Total flux spectra were
obtained at Keck using LRIS on 1998 March 5, 6, \& 27 and 1999 January 6, and
at the Lick Observatory using the Kast double spectrograph (Miller \& Stone
1993) at the Cassegrain focus of the Shane 3-m reflector in 1998 on June 18,
July 17 \& 23, and November 25 and in 1999 on January 10, March 12, and July
7. A journal of observations is given in Table 1.

One-dimensional sky-subtracted spectra were extracted optimally (Horne 1986) in
the usual manner, generally with a width of $\sim$\ 4$^{\prime\prime}$ along
the slit.  Each spectrum was then wavelength and flux-calibrated, corrected for
continuum atmospheric extinction and telluric absorption bands (Wade \& Horne
1988), and rebinned to 2 \AA\ pixel$^{-1}$.  Polarimetric analysis was
performed according to the methods outlined by Miller, Robinson, \& Goodrich
(1988) and Cohen et al. (1997).  Observations of unpolarized standard stars on
the polarimetry night showed a flat, well-behaved polarization response over
the observed wavelength range, and observations of polarized standard stars
yielded values which agreed well with the polarizations catalogued by Turnshek
et al. (1990).

\section{RESULTS AND ANALYSIS}

\subsection{The Reddening to SN 1998S}

Since reddening affects interpretation of spectropolarimetric data as well as
relative line fluxes, we consider it first.  Determining the reddening to SN
1998S was difficult since we could not rely on empirical knowledge of the
intrinsic color of similar objects to guide us, as is often done for more
normal type II and Ia events (e.g., Schmidt, Kirshner, \& Eastman 1992; Riess,
Press, \& Kirshner 1996).  Rather, we resorted to more indirect methods.

First, the reddening due to dust in the Milky Way (MW) was taken from the dust
maps of Schlegel, Finkbeiner, \& Davis (1998).  Since SN 1998S is located far
out of the Galactic plane ($b \approx 66^\circ$), this contribution was
predictably small: $E$($B-V$)$_{\rm MW}=0.02$.  Next, the reddening
contribution from the host galaxy was estimated from the rough empirical
correlations that have been derived between the equivalent width of
interstellar Na I D absorption and reddening along the lines of sight to SNe
(Barbon et al. 1990) and Galactic stars (Richmond et al. 1994; Munari \&
Zwitter 1997).  From our day 25 (after discovery) total flux spectrum, which
has both the best spectral resolution and the least contamination from
\ion{He}{1} $\lambda5876$ emission, we measured $W_\lambda {\rm (D_1,
\lambda5895.9)} = 0.37\ $\AA\ and $W_\lambda {\rm (D_2,\ \lambda5890.0)} =
0.44\ $\AA, yielding the values $E$($B-V$)$_{\rm host}$ = 0.20 (c.f., Barbon et
al. 1990), 0.19 and 0.18 (c.f., Richmond et al. 1994), and 0.23 (c.f., Munari
\& Zwitter 1997).  When added to the MW component, these values suggest $0.20
\lesssim$ $E$($B-V$)$_{\rm total}\lesssim 0.25$.  Since we are most interested
in placing an {\it upper limit} on the reddening in the spectropolarimetric
analysis (\S3.3.2), however, the fact that these relations have recently been
shown to significantly underestimate the reddening to some SNe Ia, whose
reddening is independently estimated from the SN color, makes us treat the
range of values with some skepticism (c.f., Suntzeff et al. 1999).
Nevertheless, lacking other independent estimates we will take $E$($B-V$)$\ =
0.23$ as the reddening value, and cautiously allow 0.25 to serve as the upper
bound.
 
\subsection{Spectra of SN 1998S}
\subsubsection{Early-Time Spectra}

The spectral development of SN 1998S from days three to five after
discovery is shown in Figure 2, along with probable line identifications
(c.f., Garnavich et al.  1998b).  Measured quantities at all spectral
epochs are given in Table 2.  The prominent early-time line features
closely resemble those of SN 1983K: Broad emission lines of the
hydrogen Balmer series, \ion{He}{2} $\lambda\lambda 4686, 5412$, and
\ion{C}{3}/\ion{N}{3} $\lambda 4640$ superposed on a smooth, blue
continuum (c.f., Niemela et al. 1985).  Broad features identified with
\ion{He}{1} $\lambda 6678$, \ion{He}{2} $\lambda\lambda 7593, 8237$,
\ion{C}{2} $\lambda 7100$, \ion{C}{3} $\lambda 5696$, and \ion{C}{4}
$\lambda\lambda 5801, 5812$ are also seen; the high-ionization carbon
and nitrogen lines are features commonly observed in spectra of
Wolf-Rayet stars.  The line profile of H$\alpha$ is best modeled by a
two-component system, consisting of an unresolved [full width at half
maximum (FWHM) $< 300$ \ km s$^{-1}$] narrow line centered on a broad
[full width near zero intensity (FWZI) $\gtrsim 20,000$ km s$^{-1}$],
modified Lorentzian base (see Fig. 3).  In addition, narrow,
unresolved lines of [\ion{O}{3}] $\lambda\lambda 4363, 4959, 5007$,
\ion{He}{1} $\lambda\lambda 5876, 6678, 7065$, and \ion{He}{2} $\lambda
8237$ are present.  The strength of the broad features steadily
decreased over the three observations, with the equivalent width of
H$\alpha$ dropping from 64 \AA\ to 43 \AA\ during the two-day interval.

These early emission features likely result from two distinct regions of
interaction between the expanding SN and its surrounding CSM: Narrow lines from
photoionized material in a region of undisturbed CSM exterior to the SN, and
broad lines from shocked or photoionized ejecta (e.g., Chugai \& Danziger 1994;
see also Chevalier \& Fransson 1994), or perhaps overlying stellar wind
material surrounding the expanding ejecta, a possibility considered for SN
1983K (Niemela et al. 1985; see also Grasberg 1993).  The inferred density of
the CSM is very high, with narrow [\ion{O}{3}] ($\lambda\lambda 4959,
5007$)/$\lambda 4363 \approx 1.4$ (average of rough values from the spectra
obtained on days 4 and 5), implying densities of $6.0 \times 10^6 \lesssim n_e
\lesssim 2.0 \times 10^7$ (all information on the density of the CSM presented
in this paper was obtained using the Space Telescope Science Data Analysis
System software package of IRAF\footnote{IRAF is maintained and distributed by
the Association of Universities for Research in Astronomy, under a cooperative
agreement with the National Science Foundation.}) for temperatures in the
reasonable range $30,000 > T > 15,000 $ K.  Including a correction for
reddening of $E$($B-V$) = 0.23 changes this ratio slightly, to $\sim 1.23$,
suggesting even higher densities.

While considerable theoretical effort has gone into modeling the emission lines
seen in SNe IIn, the mechanism responsible for the continuum remains largely
uncertain: Is it produced by an expanding, thermal photosphere interior to a
line-forming region, or is some other mechanism operating?  At very early
times, the main continuum opacity source at optical wavelengths in ``normal''
SN atmospheres is thought to be electron scattering, which is independent of
wavelength (e.g., Eastman et al. 1994).  Since line blanketing from metals
should not significantly alter the flux distribution at early epochs, we might
expect a young SN continuum to closely approximate a Planckian energy
distribution.  In the only previous detailed study of early-time type II
continuum shape, Clocchiatti et al. (1995) indeed found good agreement between a
cooling, reddened, blackbody, and spectra sampling the first 6 days after shock
breakout of SN 1993J, a ``type IIb'' event (c.f., Woosley et al. 1987;
Filippenko 1997).  For SNe IIn, comparisons have found early-time spectra to be
roughly consistent with a blackbody (e.g., Niemela et al. 1985; Stathakis \&
Sadler 1991; Benetti et al. 1999), although ``bluer than normal'' continua have
sometimes been noted (Schlegel 1990; Filippenko 1997).  A detailed comparison,
however, has not been done.  Since the shape and relative flux of the strong
continuum for our early spectra should be very accurate and galaxy subtraction
errors minimal (the galaxy background was only a few percent of the SN's
brightness over the spectral range covered, and less than 1\% at $\lambda <
6000 $ \AA), we have an excellent opportunity to investigate this question.

The continuum grew steadily redder over the two-day interval spanned by the
three spectral epochs, suggesting a cooling source.  To see whether it could be
modeled by a blackbody, we first assumed $E$($B-V$) = 0.23 (\S3.1) and $R_V=
3.1$ (Savage \& Mathis 1979), and fit for the best photospheric temperature,
using the reddening law of Cardelli, Clayton, \& Mathis (1989) with the
O'Donnell (1994) modification at blue wavelengths.  We assumed a pure
emission-line spectrum, and discarded fits that rose more than 2\% (the
estimated upper bound to the uncertainty in continuum placement) above the
continuum at any point.  Due to its large spectral range, we focused our
analysis on the day four spectrum.  One difficulty is that the region $\lambda
\lesssim 5178$ \AA\ is filled with overlapping line features, making the
placement of the continuum there quite uncertain.  Thus, we first fit to the
region $\lambda > 5178$ \AA, defining our continuum in ``line-free'' regions
(i.e., more than 10,000 km s$^{-1}$ away from known broad features) and
connecting adjacent regions with a smooth spline.  When this was done, the best
fit for $E$($B-V$)$ = 0.23$ and $R_V=3.1$ was found at $T = 28,000$ K (line
({\it a}) in Fig. 2), a photospheric temperature which is quite high for a SN
II more than four days after shock breakout; at a similar epoch, SN 1993J was
found to be at only $T=13,900$ K (Clocchiatti et al. 1995; for theoretical
models producing similarly low photospheric temperatures in SNe II see Eastman
et al. 1994; Grasberg \& Nadezhin 1991; Woosley et al. 1994).

Even at such a high temperature, though, the blackbody flux seems unable to
match the likely continuum flux at blue wavelengths.  One obvious solution to
consider is that the overlapping emission lines provide the false impression of
a rising blue continuum.  When the model continuum is subtracted from the
spectrum (Fig. 4a), however, the inadequacy of the fit becomes obvious; indeed,
attempts to reproduce the continuum-subtracted region $4000$ \AA\ $< \lambda <
5178$ \AA\ with line profiles modeled by the H$\alpha$ line were quite
unsuccessful.  Spectropolarimetric evidence will also be shown to argue against
this explanation (\S4.2).  Before concluding that the continuum is not
intrinsically Planckian (or, at least, possesses a significant non-blackbody
component or modification of its blackbody spectrum), though, we must consider
the possibility that our estimates for either the reddening, the extinction
law, or both are incorrect.

When we allowed the reddening to vary freely in our fits (keeping $R_V=3.1$),
we found no significant improvement, even when attempting photospheric
temperatures as high as $10^5$ K.  When we let $R_V$ vary, keeping $E$($B-V$) =
0.23, better fits were found for $R_V \gtrsim 4.0$, but required $T \gtrsim
45,000$ K.  Similarly good fits were obtained at lower photospheric
temperatures by allowing $E$($B-V$)\ to vary as well: For $20,000 < T < 30,000$
K, good fits exist for $0.08 \lesssim E(B-V) \lesssim 0.15$ with $R_V \gtrsim
4.0$; an example is shown as line ({\it b}) in Figure 2, with the
continuum-subtracted flux shown in Figure 4b.  We note that such an unusual
extinction law would not be wholly unexpected: In the MW, values as high as
$R_V = 5.6$ have been seen along lines of sight to dense molecular clouds
(e.g., the Orion Nebula; Cardelli et al. 1989), a likely star-forming
environment for a type II SN.  A reddening different than $E{\rm (}B-V{\rm )} =
0.23$ would also not be surprising, given the uncertainty in the method used to
derive that value.

We are thus faced with the conclusion that a single-temperature blackbody
cannot explain the early-time continuum seen in SN 1998S unless a high
continuum temperature and a non-standard extinction law are invoked.  Since a
non-standard extinction law has spectropolarimetric consequences, we will
return to this discussion on the early-time continuum shape in \S 4.2, after
the spectropolarimetric analysis is complete.

\subsubsection{Later-Time Spectra}

Within a week after our day five observation the broad emission lines had
nearly disappeared (Qiu et al. 1998a), similar to the behavior of SN 1983K
(Niemela et al.  1985).  Certainly by day 25, SN 1998S had entered a new phase
of spectral evolution, one no longer dominated by the symmetric emission
features so evident at early times (Fig. 5).  Rather, the optical spectrum
displayed very weak emission with broad (FWHM $\approx 3000$ km s$^{-1}$) and
narrow (FWHM $\approx 300 $ km s$^{-1}$) P-Cygni absorption.  The velocity
implied by the minima of the broad absorption troughs of hydrogen (H$\alpha$,
H$\beta$, H$\gamma$), helium (\ion{He}{1} $\lambda\lambda5876, 7065$), iron
(\ion{Fe}{2} $\lambda\lambda5018, 5169$) and, more tentatively, silicon
(\ion{Si}{2} $\lambda6350$; c.f., Garnavich et al. 1998b), scandium
(\ion{Sc}{2} $\lambda\lambda5527, 5658$), and oxygen (\ion{O}{1} $\lambda7774$)
all yield values between 4000 and 5000 km s$^{-1}$.  We note that some of the
spectral features seen in the 25 d spectrum, most notably the strong
\ion{Si}{2} ($\lambda6355$) and \ion{O}{1} ($\lambda7774$) absorption and weak
H$\alpha$ emission, resemble early-time features of SNe Ic (e.g., SN 1987M, SN
1990aa, SN 1991A; c.f., Filippenko 1992), suggesting hydrogen-deficient
expanding ejecta.  The presence of external illumination of the line-forming
region by light from circumstellar interaction might also be contributing to
the low contrast of the lines relative to the continuum (c.f., Branch et
al. 1999).  The highest velocity gas, inferred from the blue edge of the broad
absorption troughs in H$\alpha$ and \ion{Si}{2} $\lambda6355$, is at $v_{\rm
max}\approx 7000$ km s$^{-1}$.  The narrow absorption features found in the
Balmer hydrogen series and \ion{He}{1} ($\lambda\lambda5876, 6678, 7065$) lines
all indicate a blueshifted velocity of $100-200$ km s$^{-1}$ from the rest
wavelength, although blending from the associated narrow emission components
(centered on the rest wavelengths; c.f., Fig. 7) probably makes these troughs
appear more blueshifted than they would be in the absence of emission-line
contamination.  From day 108 onwards, strong, broad (FWZI $\approx 14,000$ km
s$^{-1}$ for H$\alpha$), asymmetric emission lines dominate the spectrum
(Fig. 6).  The extraordinary development of the H$\alpha$\ profile is shown in
Figure 7.  A distinct blue peak at about $-4200$ km s$^{-1}$, evident as early
as day 108, comes to dominate the profile by day 268; by day 312, a decidedly
double (or even triple) peaked profile has developed, with the blue peak
continuing to increase relative to the red.  A similar profile is seen in
H$\beta$ as well.

High density of the CSM continues to be implied at these later times, with
narrow [\ion{O}{3}] ($\lambda\lambda 4959, 5007$)/$\lambda 4363 \approx 5.8$
(average of values from days 140, 312, and 375; including a correction for
reddening of $E$($B-V$) $ = 0.23$ changes this ratio to $\sim 5.1$), suggesting
a density of $n_e \approx 1 \times 10^7\ {\rm cm^{-3}}$ for a likely
temperature of 10,000 K (c.f., Terlevich et al. 1992; Chugai 1991), consistent
with the density derived from earlier epochs.  Further evidence for high
density comes from the H$\alpha$/H$\beta$\ ratio ($\gtrsim 10$ in the day 312,
day 375, and day 494 spectra), and the appearance of \ion{O}{1} $\lambda8446$,
most evident in the day 312 and day 375 spectra, though probably blended with
\ion{Ca}{2} in the earlier spectra.  The existence of this \ion{O}{1} line and
the absence of the typically stronger \ion{O}{1} $\lambda7774$ \ line suggest
that it is produced by Ly$\beta$\ pumping, implying material optically thick in
H$\alpha$ (Grandi 1980).

Overall, the general characteristics of these spectra suggest a complex
interaction between the SN ejecta and its immediate circumstellar environment.
The emission-dominated early-time and later-time spectra most certainly result
from interaction between the ejecta and dense CSM, with the reported infrared
color excess (Garnavich et al. 1998a) and detection of a CO overabundance in
infrared spectra (Gerardy et al. 1998) likely heralding the formation of dust
in the ejecta.  The increasing strength of the blue side of the line profiles
relative to the red (seen in both H$\alpha$ and H$\beta$ in Fig. 7; see also
Table 2) is also consistent with dust formation, since red photons, produced on
the far, receding side of the expanding ejecta, would suffer more extinction
from dust than blue ones due to their greater path length through the dusty
ejecta.  We will return to the geometry and evolution of the 1998S system
(\S4.3) after the implications of the early-time spectropolarimetric
observation have been discussed.

\subsection{Spectropolarimetry of SN 1998S}
\subsubsection{Total Polarization}

The polarization data obtained for SN 1998S on 1998 March 7, five days after
discovery, are shown in Figure 8. A few things are readily apparent: (1) the
continuum polarization is roughly constant across the spectrum, at $p \approx
2\%$; (2) decreases in the polarization are seen at the location of emission
features in the flux spectrum; (3) the polarization change across all of the
broad lines occurs mainly in the $u$ direction, thus forming straight
line-segments which all point in the same general direction in the $q$-$u$
plane (see Fig. 9); and (4) sharp polarization features distinct from the broad
lines are seen in the narrow, unresolved cores of some of the strongest lines
(most notably H$\alpha$, but also evident in H$\beta$, \ion{He}{1} $\lambda
5876$, and \ion{He}{2} $\lambda 4686$). Although most easily observed in the
$q$ direction (since the broad component contributes little there), the narrow
features exist in $u$ as well (most easily seen in the \ion{He}{1} $\lambda
5876$ line since there was little, if any, broad-line emission).  Such narrow
features have not previously been detected in SN polarization studies, and
multiple extractions using both the standard and optimal algorithms with
different ``object'' and ``sky'' regions show that they are real, and not an
artifact of a particular extraction choice.  Careful examination of the
two-dimensional data frames also revealed no significant \ion{H}{2} region
emission extending beyond the spatial extent of the SN, and the small
possibility that the effect is due to contamination of the SN light by
foreground \ion{H}{2} region light directly along the line of sight to the SN
is doubtful, since the effect exists in the \ion{He}{2} $\lambda 4686$ narrow
line, an unlikely species for \ion{H}{2} region light.

Taken at face value, the continuum polarization of $p\approx 2\%$ implies an
asphericity of $\gtrsim 30\%$ for the continuum scattering atmosphere of SN 1998S
(c.f., H\"{o}flich 1991).  Before interpreting the results further, however, we
must grapple with that elusive gremlin of polarimetric data, the ISP.

\subsubsection{Setting Limits on the ISP}

A problem which plagues interpretation of all SN polarization measurements is
proper removal of the ISP: Aspherical dust grains preferentially aligned with
the magnetic field of either the host or MW galaxy can contribute a
polarization signal which dwarfs that of the SN.  Fortunately, the ISP of the
MW has been well studied and shown to be a smoothly varying function of
wavelength, and constant with time (e.g., Serkowski, Mathewson, \& Ford 1975;
Whittet \& van Breda 1978).  If similar ISP characteristics are present in
other galaxies (c.f., Jones 1989; Scarrott, Rolph, \& Semple 1990) then
intrinsic SN polarization is implied by any of the following: (1) temporal
changes in overall polarization level (requires multi-epoch observations); (2)
distinct spectral polarization features (requires good resolution
spectropolarimetry); or (3) continuum polarization characteristics clearly
different from the known form produced by interstellar dust (i.e., the
``Serkowski Law''; Serkowski 1973; Wilking, Lebofsky, \& Rieke 1982).  Relying
on these criteria, we find clear evidence for polarization intrinsic to SN
1998S since it definitely satisfies criterion 2, and probably 3 as well.  To
progress further along quantitative lines, we now attempt to disentangle the SN
polarization from that contributed by ISP.

Unfortunately, while subtracting the polarization in $q$-$u$ space of a
Galactic star along the line of sight can help remove Galactic ISP (e.g., Tran
1995), there is no similarly straightforward way to remove host galaxy
contamination, an especially problematic situation for core-collapse objects
which are frequently embedded in dusty regions.  We can, however, place rough
upper limits on the ISP, and interpret our results in terms of these limits.
The essential idea is that our estimated reddening limit can be translated into
a maximum ISP through the empirical formula $P_{\rm max}=9$$E$($B-V$)\
determined from observations of MW stars (Serkowski et al. 1975).  If we assume
this relation holds for all galaxies (c.f., Jones, Klebe, \& Dickey 1992), then
our limiting reddening value of $E$($B-V$)$\ <\ 0.25$ leads directly to ${\rm
ISP}_{\rm max} = 2.25\%$.  Note that the majority of this limiting value comes
from the host galaxy, consistent with polarization measurements of Galactic
stars near the line of sight to SN 1998S: Of the nine stars within 10$^\circ$
of SN 1998S contained in the catalogue by Mathewson et al. (1978), all have
$p\leq0.06\%$, although only one (HD98839, $p = 0.02\%$) lies sufficiently far
away ($\sim 250$ pc) to fully sample the ISP through the Galactic plane.

Since ISP is nearly additive in Stokes $q$ and $u$ when the ISP is small,
effectively defining an origin in the $q$-$u$ plane for calculating the
intrinsic polarization (i.e., the distance from the ISP to the data points),
the clearest presentation of polarization data is in the $q$-$u$ plane.  In
Figure 9 we show the $q$ and $u$ data (rebinned to 10 \AA\ pixel$^{-1}$ to reduce
noise), along with the limits on the ISP determined earlier.  To illustrate the
dramatic effects that can result from different ISP values, we chose four ISP
values (indicated in Fig. 9; detailed descriptions of how points ($a$) and
($b$) were chosen are given in \S 3.3.3 and \S 3.3.4) and show the resulting
``intrinsic'' polarization in Figure 10.  As can be seen, line polarization dips
can become peaks or even disappear completely, and the overall level of
continuum polarization can either decrease {\it or increase} upon removal of
the ISP.  Clearly, any interpretation of SN polarization must necessarily be
bounded by the uncertainty in the ISP.

For SN 1998S, however, we may be able to place additional constraints on the
ISP from physical considerations.  The polarization seen in Figure 8b broadly
suggests the physical picture of a continuum source surrounded by an aspherical
electron-scattering atmosphere with an overlying emission-line region.  Such a
situation would explain the flat continuum polarization (due to the
wavelength-independent nature of electron scattering) as well as the reduced
polarization in line features (from the dilution of polarized continuum light
by unpolarized line recombination emission).  In detail, however, the existence
of different polarization properties for the broad and narrow lines mentioned
in \S 3.3.1 can be interpreted in several ways, depending on which sets of
lines are considered to be polarized.  (The different directions taken in the
$q$-$u$ plane by the broad and narrow lines rule out the possibility of both
components being unpolarized.)

Physically, the answer is determined by whether sufficient free electrons exist
in the environment of the line's production to scatter, and therefore polarize,
the line's light.  Since broad, unblended emission features have traditionally
been treated as intrinsically unpolarized in SN studies (c.f., Tran et
al. 1997; Trammell et al. 1993; H\"{o}flich et al. 1996; Jeffery 1991b; Wang et
al. 1996), we follow this path first, and examine its implications for the
geometry of SN 1998S.  Then, in \S 3.3.4 and \S 3.3.5, we address the question
of how different assumptions (e.g., narrow lines unpolarized and broad lines
polarized, or both broad and narrow lines polarized) affect the implied
geometry.

Unpolarized broad-line photons diluting polarized continuum light
produce predicted polarization characteristics which can be compared
with what is observed. If this model were exactly correct the
polarized flux (i.e., ``Stokes Flux,'' $p \times f$) would simply be a
noisy version of the continuum with no broad-line features evident,
since the amount an unpolarized emission line reduces the polarization
level at a given wavelength would be the same as the amount the flux
would increase relative to the continuum at that same wavelength.
Assuming zero ISP, the Stokes flux nearly realizes this prediction
(Fig. 11a), following the continuum shape throughout the spectrum,
with only small decreases evident at H$\alpha$, H$\beta$, and
\ion{He}{2} $\lambda 4686$, and a slight increase at
\ion{N}{3}/\ion{C}{3} $\lambda 4640$.  The obvious question, then, is
whether a unique ISP exists which can completely remove these line
features; in addition to bolstering confidence in the model, the
existence of such a value would also yield an estimate of the ISP.  To
estimate the ISP, we follow the method of Tran et al. (1997), in which
the $q$ and $u$ ISP values (i.e., $q_i$, $u_i$) that individually make
a line's feature in the ``Stokes parameter fluxes'' ($q\times
f_\lambda$ and $u\times f_\lambda$; see Figs. 11c and 11d) disappear
are found.  This is accomplished by comparing the continuum-subtracted
Stokes parameter fluxes in a line with the continuum-subtracted total
flux spectrum in the line, yielding
\begin{eqnarray}
q_i = q\times f_\lambda{\rm (cs)}/f_\lambda {\rm (cs),}\nonumber\\
u_i = u\times f_\lambda{\rm (cs)}/f_\lambda {\rm (cs),}\nonumber
\end{eqnarray}
\noindent where ``cs'' denotes continuum-subtracted line fluxes; see Tran et
al. (1997) for a thorough discussion of this method.

The major source of uncertainty is accurate continuum placement in both the
total and Stokes parameter fluxes.  Fortunately, since our total flux
spectrum's continuum is very well defined, placement there is straightforward;
the main source of error comes from estimating the continuum in the Stokes
parameter fluxes, due to their inherently noisy nature.  When we apply this
technique to H$\alpha$ (it is the best line since it is not blended with other
strong features), we derive $q_i=-0.9 \pm 0.3$, $u_i=1.4 \pm 0.4$ at $\lambda
\approx 6563 $ \AA, with errors derived from the estimated uncertainty in
continuum placement.  For a Serkowski law with the wavelength of maximum
polarization occurring at $\lambda_{\rm max} = 5600 $ \AA\ this becomes $p_i=1.7
\pm 0.4\% {\rm \ at\ PA} = 62^\circ \pm 6^\circ$, illustrated as point ($a$) in
Figure 9.  Adopting this ISP produces the ``intrinsic'' polarization shown in
Figure 10a, and the Stokes flux, Stokes parameter fluxes, and PA shown in
Figure 12.  Notice that with the removal of this ISP the intrinsic continuum
polarization has increased to $p \approx$\ 3\% (at PA $\approx 134^\circ$),
implying a global asphericity of at least 45\% according to the models of
H\"{o}flich (1991).

The value for this ISP ($p_i=1.7 \pm 0.4\%$ at PA = $62^\circ \pm 6^\circ$) is
encouraging on several fronts.  First, it lies within the upper limit of 2.25\%
derived from reddening considerations (\S 3.3.2).  Second, in addition to
removing the H$\alpha$ feature in polarized flux, it effectively removes all
but one of the other broad-line features (Fig. 12a).  A natural explanation for
the different behavior of the \ion{C}{3}/\ion{N}{3} $\lambda 4640$ line, which
is seen to increase in total Stokes' flux above the continuum value, is that it
is produced in a slightly different physical location, perhaps closer to or
even within the primary electron-scattering region of the continuum atmosphere.
This would allow line photons to become polarized as they scatter their way out
of the atmosphere; interestingly, a similar ionization stratification has been
noted in Wolf-Rayet stars' atmospheres (Harries et al.  1999).  Third, recent
polarization studies of edge-on spiral galaxies show polarization vectors
closely aligned parallel to the galaxy's disk (e.g., Jones 1997; Scarrott et
al. 1990), and the nearly edge-on host of SN 1998S ($i < 14^\circ$; c.f., Frei
et al. 1996) has a PA of $35^\circ$ on the sky, reasonably close to our
measured ISP PA.  Finally, adopting this ISP has made the rotations in
polarization PA originally seen across the broad emission lines (Fig. 11b)
essentially disappear (Fig. 12b), suggesting that even if the broad-line
photons are themselves slightly polarized by scattering, the geometry of their
scattering environment is at least the same as that of the continuum.  The
rotation seen in the narrow lines' polarization PA after removal of this ISP,
however, implies scattering of narrow-line photons from material with a
different geometric distribution than the continuum in this model.

Two additional points from this investigation are worth mentioning.  First, the
inferred value of the ISP really places a lower bound on the intrinsic SN
polarization for this model: If the H$\alpha$ line photons are themselves
slightly polarized, then the true ISP value is actually farther away from the
continuum (i.e., $u_i$ increases), thereby {\it increasing} the implied SN
polarization and corresponding asphericity.  Second, the straight line-segments
traced in $q$-$u$ space by the broad-line features do not necessarily imply
axisymmetry since the completely unpolarized line photons assumed by this model
will produce lines (as opposed to loops) in $q$-$u$ for {\it any} geometry of
the line-forming region.

\subsubsection{Interpretation II:  Broad Lines Polarized, Narrow Lines
Unpolarized}

Overall, the assumption of unpolarized broad lines with polarized narrow lines
described in the previous section appears to have served us well, producing a
believable ISP and a reasonably straightforward physical interpretation.
However, we must also consider the different, though plausible, assumption that
the broad lines are polarized and the narrow lines are unpolarized or, 
equivalently, are scattered by a spherical distribution of electrons.

We can again use the method of Tran et al. (1997) to determine the preferred
ISP under these new assumptions.  That is, we now seek to eliminate the {\it
narrow} features in the Stokes parameter fluxes, just as we removed the broad
features earlier.  This is done primarily with H$\alpha$, though the other
narrow lines are used as well to help determine $u_i$, due to the difficulty of
separating the narrow from the broad component in the $u \times f_\lambda$
H$\alpha$ profile.  We derive $q_i=2.0 \pm 0.5 \%, u_i=1.5 \pm 0.6\%$, or
$p_i=2.5 \pm 0.5\%$ at PA = $18^\circ \pm 6 ^\circ$, with errors again
estimated mainly by the uncertainty in the continuum placement around the
narrow features.  Note that this inferred ISP strains the limit imposed by
reddening considerations, shown as point ($b$) in Figure 9.

As expected, this ISP successfully removes the narrow features in the Stokes
parameter fluxes and the PA rotations across them (Fig. 13).  By doing so,
however, it has introduced polarization and PA rotation to the broad-line
features.  The physical inferences that are drawn from this model are thus
quite different from those found before.  First, the resulting ``intrinsic''
continuum polarization of $p \approx 4.3\%$ shown in Figure 10b implies an
asphericity of at least 60\% (c.f., H\"{o}flich 1991).  Second, the rotation of
the polarization PA across the broad-line features necessitates a different,
though axisymmetric (straight lines traced in the $q$-$u$ plane), scattering
geometry for the broad lines than for the continuum region.  

\subsubsection{Interpretation III:  Both Broad and Narrow Lines Polarized}

The final possibility to consider is that {\it both} the broad and narrow
components are polarized, and have different scattering geometries than the
continuum region.  With this assumption, we are left unable to determine the
ISP, and can draw only limited conclusions: 1) the geometry of the broad line's
scattering environment is axisymmetric; 2) the geometries of the broad and
narrow-line scattering environments are different; and 3) the continuum
polarization $0\% \leq p \leq 4.2\%$, from the limits imposed by the reddening.
Example ISP values allowed under this interpretation are shown in Figure 9 as
points ($c$) and ($d$), with the resulting intrinsic polarizations given in
Figures 10c and 10d.  We note that assuming zero ISP (Figs. 8 and 11) also
falls under this interpretation.

\section{DISCUSSION} 
\subsection{Polarization}

Spectropolarimetry is notoriously difficult to interpret, given the
uncertainties in both the interstellar contribution and the models of the
immediate scattering environment for the object itself.  As we have seen, our
single spectropolarimetric observation of SN 1998S is quite open to competing
interpretations, resulting from different assumptions made for the polarization
properties of the broad and narrow emission lines.  Which assumptions are most
reasonable?  To be sure, additional early-time spectropolarimetric epochs would
have helped lay many choices to rest.  For instance, if the continuum
polarization varied and either the broad-line or narrow-line features pointed
to a consistent position in the $q$-$u$ plane, a depolarizing set of lines with
geometry consistent with the continuum would be implied, allowing a unique ISP
to be derived; see Harries et al. (1999) for an application of this elegant
technique.  Lacking such information, however, we must speculate largely from
considerations of physical reasonableness.

Models of SNe interacting with a dense CSM generally envisage a double
shock-wave structure near the ejecta-CSM interface, with one shock propagating
outward into the wind material (outer shock) and the other moving inward
through the ejecta (inner shock; c.f., Chevalier 1982).  If the inner shock is
radiative, then a dense shell of shocked ejecta will form behind it at the
contact discontinuity between the inner and outer shocks.  It is this thin,
dense shell that has been proposed as the likely source of the broad-line
emission features, energized by either the reprocessing of X-rays emitted by
the shock waves or through thermal conductivity with the hot, shocked wind
plasma (Chugai 1997b).  The densest part of this shell may also be responsible
for the continuum production (Chugai 1991).  The narrow lines, on the other
hand, are thought to come from photoionized stellar wind material in regions as
yet undisturbed by the expanding ejecta.  The narrow lines thus serve as a sort
of precursor, revealing information about the CSM the ejecta will encounter in
the future.

From the broad-line features seen in the total flux spectra of SN 1998S, we can
identify at least two epochs of ejecta-CSM interaction: The first, lasting
$\lesssim 12$ days after discovery, with material in the immediate vicinity of
the progenitor, and the second, evident from day 108 onwards though presumably
starting somewhat before this, with a distribution of material farther away.
This second region of material is likely the gas mainly responsible for the
narrow lines, suggesting that $n_e \approx 10^7$ cm$^{-3}$ there, a density
which could plausibly produce some optical depth to electron scattering.
Further, the blue and red peaks of the broad emission profiles evident in
later-time spectra may indicate that this CSM is not spherically distributed,
but is instead concentrated in a ring or disk (see, e.g., Warner 1995,
pp. 93-98).\footnote{For an alternate broad-line production scenario which does
not require an asymmetric CSM yet still produces a double-humped profile, see
Chugai (1991).}  Evidence from the total flux spectra can thus naturally create
the physical picture of a dense, asymmetric scattering environment for the
narrow lines.

On the other hand, the model of unpolarized narrow lines seems to us more
difficult to maintain.  Such an assumption requires an ISP that pushes the
allowed limit and a broad-line scattering geometry that is both axisymmetric
and different than the continuum region's geometry, a notion that runs counter
to their probable similar production locations.  This last issue also argues
against the picture of the narrow and broad lines both being polarized, since
this too necessitates different geometries for their scattering environments.

Though impossible to know for sure, we are therefore led to prefer the
``conventional'' interpretation of depolarizing broad lines surrounding a
polarized continuum.  If correct, then the intrinsic SN polarization of $p
\approx 3\%$ is the highest SN polarization thus far observed.  Interestingly,
the previous record of $p \approx 1.5\%$ was shared by SN 1994Y and SN 1993J
(Wang et al. 1996), both of which are believed to have experienced significant
mass-loss prior to explosion.  Indeed, though still limited by small numbers,
there seems to be a trend of increasing SN polarization with decreasing H or He
envelope mass (Reddy, H\"{o}flich, \& Wheeler 1999).

\subsection{Early-Time Continuum}
We now return to our discussion of the early-time continuum shape, viewed in
light of our preferred interpretation of the spectropolarimetry.  One
possibility previously considered was that the excess blue flux could be
explained by a series of overlapping emission lines.  If such lines had
polarization characteristics similar to the other emission lines (i.e., they
contribute diluting, unpolarized light), we would expect the overall
polarization level to decrease towards blue wavelengths, an effect not seen to
any significant degree (Fig. 10a).  A second idea proposed was that the blue
excess was partly an artifact of an unusual extinction law, with $R_V \gtrsim
4.0$.  Since lines of sight to reddened Galactic stars obey the relation $R_V=
5.6\lambda_{\rm max}$(\AA)$/10,000$ (Whittet \& van Breda 1978), $R_V \gtrsim
4.0$ leads to the expectation of $\lambda_{\rm max} \gtrsim 7100$ \AA.
However, for the preferred ISP we find that any $\lambda_{\rm max} \gtrsim
6100$ \AA\ results in a continuum polarization PA that changes significantly
(i.e., $\gtrsim 1^\circ$) across the spectrum, inconsistent with the
wavelength-independent geometry assumed in the model.  Placing faith in this
model, then, necessarily argues for a more normal $R_V\approx 3.1$ and against
either an unusual extinction law or overlapping emission lines as the cause of
the blue continuum shape.\footnote{We note that subtle combinations of the
parameters, however, cannot be absolutely ruled out by the polarimetry alone,
due mainly to its limited wavelength coverage.  For instance, a slightly
unusual extinction law of $R_V \approx 3.4$ ($\lambda_{\rm max}=6000$ \AA)
provides a PA which is uniform to within our uncertainty ($\sim 1 ^\circ$).
When a Serkowski law with $\lambda_{\rm max}=6000$ \AA\ at our preferred ISP is
removed from the polarization data, a small drop in Stokes flux occurs at blue
wavelengths.  Since the Stokes flux reveals the true continuum shape, this
change permits blackbody fits at lower temperatures, though for $E$($B-V$)$ =
0.23$ temperatures in excess of 40,000 K are still required.  To get $T <
15,000$ K requires $E(B-V) < 0.03$, quite different from the reddening
estimated in \S 3.1.}

One final possibility to consider for the blue continuum is that light
scattered by dust in the circumstellar nebula surrounding SN 1998S is producing
a ``light echo,'' resulting in SN light from an earlier epoch contributing to
the observed spectrum (e.g., Chevalier 1986; see also Lawrence, Crotts, \&
Gilmozzi 1998).  Not only is the scattered light from a hotter (i.e., bluer)
epoch, but it should further increase at shorter wavelengths since dust grains
with size typical for the interstellar medium ($\lesssim 0.25\ \mu{\rm m}$)
have a scattering efficiency that rises sharply with decreasing wavelength
(c.f., Spitzer 1978).  Could this explain the unusually blue continuum seen in
the early-time spectra of SN 1998S and SNe IIn in general?  Though impossible
to completely discount given our limited data, evidence can certainly be found
to counter this model.  First, the extreme youth of SN 1998S coupled with the
likely destruction of dust grains by the intense UV flash within a radius of
about seven light-days (Wang \& Wheeler 1996) greatly limit the region of space
that could contribute to such an echo.  Second, attempts to fit the overall
light curves of SNe IIn with the simple light-echo model have been unsuccessful
(Roscherr \& Schaefer 1998), although contributions at the $\sim 5\%$ level
needed here probably cannot be ruled out.  Finally, if the proposed dust exists
in clumps or is in any way asymmetrically distributed, then even a small amount
of scattering should produce measurable polarization in the received SN light,
since dust scattering is very efficient at polarizing light.  Although
computing a detailed light-echo polarization model is beyond the scope of this
study (see Wang \& Wheeler (1996) for such modeling in the case of SN 1987A),
we note that simple reflection (i.e., single scattering) off grains of typical
size should produce polarization which increases at blue wavelengths due to the
greater scattering efficiency there (e.g., Miller \& Goodrich 1990; Miller,
Goodrich, \& Mathews 1991; see, however, Kartje 1995, for a discussion of how
multiple scatters could alter this), a trend inconsistent with the flat
continuum polarization observed here in SN 1998S.

Since ``extrinsic'' effects seem unlikely, we propose that the continuum, while
predominantly blackbody in nature, has either an additional component or a
modification of its blackbody spectrum producing a color temperature which
increases with decreasing wavelength.  Since unusually blue continua have been
observed in previous early-time SNe IIn spectra, and a blue excess was not
detected in similarly early-time spectra of SN 1993J (Clocchiatti et al. 1995),
it may be that this characteristic is unique to SNe IIn, a situation that would
favor mechanisms requiring dense circumstellar environments over sources that
are common to all SNe.  Obvious possibilities include free-free emission and
Comptonization of photospheric photons by hot ($T\approx 10^9 {\rm\ K}$)
electrons in the shocked gas, sources which have been invoked to explain flux
excesses seen in the far ultraviolet ($\lambda < 1500\ $\AA) in previous SNe
interacting with dense circumstellar environments (e.g., 1979C; Fransson 1982),
and even SN 1998S itself later in its evolution (Lentz et al. 1998).  The
amount these processes are predicted to contribute to the total flux spectrum
at the optical wavelengths considered here, however, is probably $\lesssim 1\%$
(c.f., Fransson 1982), although a detailed model with conditions specific to SN
1998S has not been done.  A more likely contributor may be the wavelength
dependence of the continuum absorptive opacity, which increases with wavelength
for both bound-free (i.e., the Paschen continuum) and free-free processes
through most of the spectral range observed (see, e.g., Eastman et al. 1994).
Significant optical depth for these processes is possible in the cool, dense
shell ($T \approx 10^4 {\rm\ K}$, $n\approx 10^{13} {\rm\ cm}^{-3}$; c.f.,
Chugai 1991) of shocked ejecta where the continuum is probably produced.  Such
continuous opacity could explain both the smooth appearance and apparent blue
excess seen.

\subsection{Geometry of SN 1998S and its CSM}

By combining our preferred interpretation of the spectropolarimetry (\S 4.1)
with information derived from the total flux spectra (\S 3.2) it is possible to
construct a rough evolutionary scenario for SN 1998S.  The main observational
points are: (1) the early-time continuum is produced within a highly
aspherical ($>45\%$) scattering region; (2) strong, broad (half width near zero
intensity (HWZI) $\gtrsim 10,000\ {\rm km\ s}^{-1}$), symmetric emission
features seen early disappear by day 12; (3) at day 25 the spectrum has
features in common with early-time spectra of stripped-envelope SNe; (4) at day
108 the spectrum exhibits a broad, asymmetric H$\alpha$ emission line; and (5)
between day 140 and day 268 the H$\alpha$ profile becomes double (or even
triple) peaked.

We start by setting approximate size constraints on the CSM.  For the purposes
of this simple model, we assume that the explosion occurred on the
discovery date, and that the presence or absence of emission-dominated
line profiles signifies interaction, or lack thereof (respectively), between
the fastest moving ejecta (10,000 km s$^{-1}$) and the CSM.  This allows us to
identify two distinct regions of circumstellar material, one in the immediate
vicinity of the progenitor (``inner CSM'') and one farther away (``outer
CSM'').  The rather abrupt disappearance of the broad emission features by
about day 12 suggests that the inner CSM has been overrun by the expanding
ejecta, restricting its size to $\sim 70$ AU.  From the reemergence of broad
emission by day 108 we derive an inner boundary of $< 620$ AU for the outer
CSM.  Since this interaction continues through the day 494 spectrum, the outer
CSM region must extend out to $\gtrsim 3000$ AU.

These distinct phases of interaction with CSM imply episodic mass-loss of the
progenitor, with the most recent episode being terminated by the explosion
itself.  The circumstellar gas producing the narrow P-Cygni absorption in the
25 d spectrum ($v_{\rm trough} \approx\ 150\ {\rm km\ s}^{-1}$) is likely the
same gas producing the narrow emission ($v_{\rm unresolved} < 300\ {\rm km\
s}^{-1}$).  Due to the large optical depth of the lines, the shallow density
gradient of the circumstellar wind, and contamination from the narrow emission
components, the density peak in the CSM producing the blueshifted narrow
absorption is probably significantly slower than $150\ {\rm km\ s}^{-1}$ (c.f.,
Jeffery \& Branch 1990).  In order to set the time evolution of the
progenitor's mass-loss, we will assume that the circumstellar gas (both outer
and inner CSM) is traveling at 50 ${\rm km\ s}^{-1}$.  With this approximation,
the boundaries previously set on the outer CSM imply a mass-loss episode which
began more than 300 years before the explosion and terminated less than 60
years before core collapse.  If we assume that the more recent mass-loss
episode is characterized by a similar velocity, then we conclude that it began
only seven years prior to core collapse.

These considerations lead to the following evolutionary scenario.  The
progenitor star of SN 1998S underwent a significant mass-loss episode
terminating less than 60 years prior to the explosion, having lasted more than
240 years.  This mass-loss may have been concentrated in a ring or disk early
on, perhaps becoming more spherical near the end.  A second mass-loss episode
began only seven years before core collapse, and continued up to the time of the
explosion.  This mass-loss may have stripped much of the remaining hydrogen
envelope from the progenitor and deposited it in the dense CSM.  When the star
exploded, it immediately interacted with the most recently lost material,
completely engulfing it within 12 days.  For a brief time ($\sim$ few weeks),
the ejecta expanded without significant circumstellar interaction, revealing
spectral features similar to other stripped-envelope events.\footnote{The lack
of a strong \ion{Ca}{2} near-IR triplet in the 25 d spectrum is also consistent
with this identification.  Filippenko (1992) found H$\alpha$ absorption
strength to be inversely correlated with the strength of the near-IR
\ion{Ca}{2} lines, perhaps indicative of the relative concentration of these
elements in the SN envelope.  The Balmer hydrogen profiles seen in SN 1998S,
while very weak compared with normal type II events, are quite strong relative
to the SNe Ic in Filippenko's study.}  Interaction with another region of dense
CSM began sometime before day 108 and continues through day 494, the last epoch
studied here.

To help fix ideas, a diagram illustrating the proposed geometry of SN 1998S and
its CSM five days after discovery (concurrent with the spectropolarimetric
observation) is shown in Figure 14.  In order to derive the absolute
orientation in the plane of the sky for the continuum, broad-line, and
narrow-line production regions, we took our preferred ISP ($q_i=-0.9, u_i=1.4 $)
and assumed oblate spheroids for the broad-line and continuum production
regions.  The narrow-line region was assumed to be concentrated in a disk of
CSM exterior to the expanding ejecta.  For the continuum, a PA $\approx
135^\circ$ (Fig. 12b) implies an oblate spheroid with its long axis oriented at
PA $\approx 45^\circ$.  The broad-line region, assumed to produce unpolarized
light in this model, necessarily has an undetermined geometry.  However, the
suggestion that the broad \ion{C}{3}/\ion{N}{3} ($\lambda 4640$) line is in
fact slightly polarized with PA similar to the continuum (\S 3.3.3) argues
for a common morphology for the two regions; we therefore represent the
broad-line region as having the same geometry as the continuum region, though
exterior to it.  The narrow-line production region has PA $\approx 0^\circ$
(see Fig. 9; the relative position of point ``b'' to point ``a'' determines the
PA of the narrow-line region), implying a scattering disk oriented at
$90^\circ$ (c.f., Wood et al. 1996), arbitrarily shown edge-on in the figure.

We note that the ring-like geometry inferred for the CSM of SN 1998S is quite
similar to that seen directly in SN 1987A (c.f., Crotts \& Heathcote 1999),
except much denser and closer to the progenitor in SN 1998S (inner ring radius
of $\lesssim 1500$ AU compared with $\sim$ 40,000 AU for SN 1987A).  Although
the agent ultimately responsible for such asymmetric mass-loss is unknown, it
is tempting to postulate that it resulted from the interaction of the
progenitor of SN 1998S with a companion star, similar to some models of SN
1987A (e.g., Soker 1999b).  It might prove fruitful to model in detail the
evolutionary scenario for SNe IIn suggested by Nomoto et al. (1995), in which
two stars merge shortly before explosion and undergo asymmetric mass-loss
through formation of a common envelope, to see if it can explain the observed
properties of SN 1998S presented here.

\section{CONCLUSION}

In this paper we have reported on an early-time spectropolarimetric observation
of SN 1998S and total flux spectra sampling the first 494 days after its
discovery.  Our main results are as follows.
\begin{enumerate}

\item The CSM surrounding SN 1998S has $n_e \gtrsim 10^7 {\rm cm}^{-3}$, and
may be asymmetrically distributed, perhaps with a disk-like or ring-like
morphology, generically similar to that seen directly in SN 1987A.

\item Uncorrected for ISP, the early-time continuum exhibits a flat
polarization spectrum with $p \approx 2\%$, and sharp decreases in
polarization at the location of both the broad and narrow emission
lines seen in the total flux spectrum.

\item The polarization features associated with the broad lines trace straight
line-segments in the $q$-$u$ plane, and all point in the same direction.

\item The polarization features associated with the narrow lines have different
properties than the broad lines, eliminating the possibility that both sets of
lines are unpolarized.

\item The early-time continuum has a color temperature that increases with
decreasing wavelength.

\end{enumerate}

We advance the following basic hypothesis.  The progenitor of SN 1998S was a
massive star stripped of most of its hydrogen envelope by at least two strong
mass-loss episodes, the first concentrated in a ring or disk and the last still
occurring when the explosion took place.  The early-time spectrum was dominated
by the intense interaction between the SN ejecta and an inner region of
circumstellar material surrounding the progenitor.  In our spectropolarimetric
observation five days after discovery, unpolarized broad-line recombination
emission dilutes the light from a highly polarized, aspherical continuum
region.  The continuum is likely produced in a dense shell of ejecta formed
near the ejecta-CSM interface; significant optical depth to the Paschen
continuum and free-free absorption in this shell may be responsible for its
unusually blue color.  Narrow-line emission is coming from, and being scattered
by, dense, asymmetrically distributed CSM exterior to the expanding ejecta.
When modeled in this way, we determine the ISP to be $p_i=1.7 \pm 0.4\%$ at PA
$\approx 62^\circ \pm 6^\circ$ (point($a$) in Fig. 9), resulting in an
intrinsic polarization for SN 1998S of $p \approx$\ 3\% at PA $\approx
134^\circ$ (Fig. 10a and Fig. 12) and implying a global asphericity of at least
45\% from the models of H\"{o}flich (1991).

Determining whether the inferred asphericity results primarily from an
asymmetrically distributed CSM or an aspherical explosion is difficult.  The
symmetry seen in the broad-line profiles (c.f., Fig. 3) might seem to indicate
a substantially spherical line-forming region in the CSM, thus favoring an
aspherical explosion as the cause of the polarization.  However, absorption
effects are known to wash out spectral signatures of asphericity in supernova
atmospheres (c.f., H\"{o}flich et al. 1996), rendering this diagnostic
inconclusive without detailed modeling.  It is thus difficult to eliminate
either an asymmetrically distributed CSM or an aspherical explosion as the
cause of the polarization, and both may actually play a role.  Future
spectropolarimetric observations of core-collapse SNe not interacting with
dense CSM (i.e., SNe II-P) should help shed light on this issue.

The conclusion that SN 1998S has a high intrinsic polarization adds to the
mounting evidence that many SNe are aspherical at some level.  While this
realization may aid in the resolution of some mysteries (i.e., ``kick''
velocities imparted to neutron stars), it certainly creates others, the most
crucial of which is the unknown agent responsible for the asphericity: Is the
explosion mechanism itself asymmetric?  Can progenitor star rotation produce
such asphericities (c.f., Fryer \& Heger 1999)?  Are companions involved?  How
well constrained are such ``extrinsic'' effects as scattering from an
asymmetrically distributed CSM (or dust) as the cause for the polarization?
Detailed analysis of the spectropolarimetry of nearby SNe (preferably at
multiple epochs), together with additional theoretical calculations, should
help bring these issues into sharper focus in the next few years.

\acknowledgments

We thank Peter H\"{o}flich, Lifan Wang, David Jeffery, and Joe Shields for
useful discussions and comments, and the staff at both the Keck and Lick
Observatories for providing excellent observing assistance.  The Keck LRIS and
Lick Kast engineering teams are especially acknowledged for their design and
construction of efficient, reliable spectrographs.  We are grateful to Ed
Moran, Weidong Li, Adam Riess, Ryan Chornock, Maryam Modjaz, and Andrea Gilbert
for help with the observations and data reduction.  Some of the data presented
herein were obtained at the W.M. Keck Observatory, which is operated as a
scientific partnership among the California Institute of Technology, the
University of California, and the National Aeronautics and Space
Administration.  The Observatory was made possible by the generous financial
support of the W.M. Keck Foundation.  This research has made use of the
NASA/IPAC Extragalactic Database (NED), which is operated by the Jet Propulsion
Laboratory, California Institute of Technology, under contract with NASA.  Our
work was partially funded by NASA through grants GO-7434 and GO-8243 from the
Space Telescope Science Institute, which is operated by AURA, Inc., under NASA
contract NAS 5-26555.  We also acknowledge NSF grant AST-9417213, as well as
support from the Sylvia and Jim Katzman Foundation.

\newpage
\begin{figure}[ht!]
\begin{center}

 \scalebox{.90}{
	\includegraphics{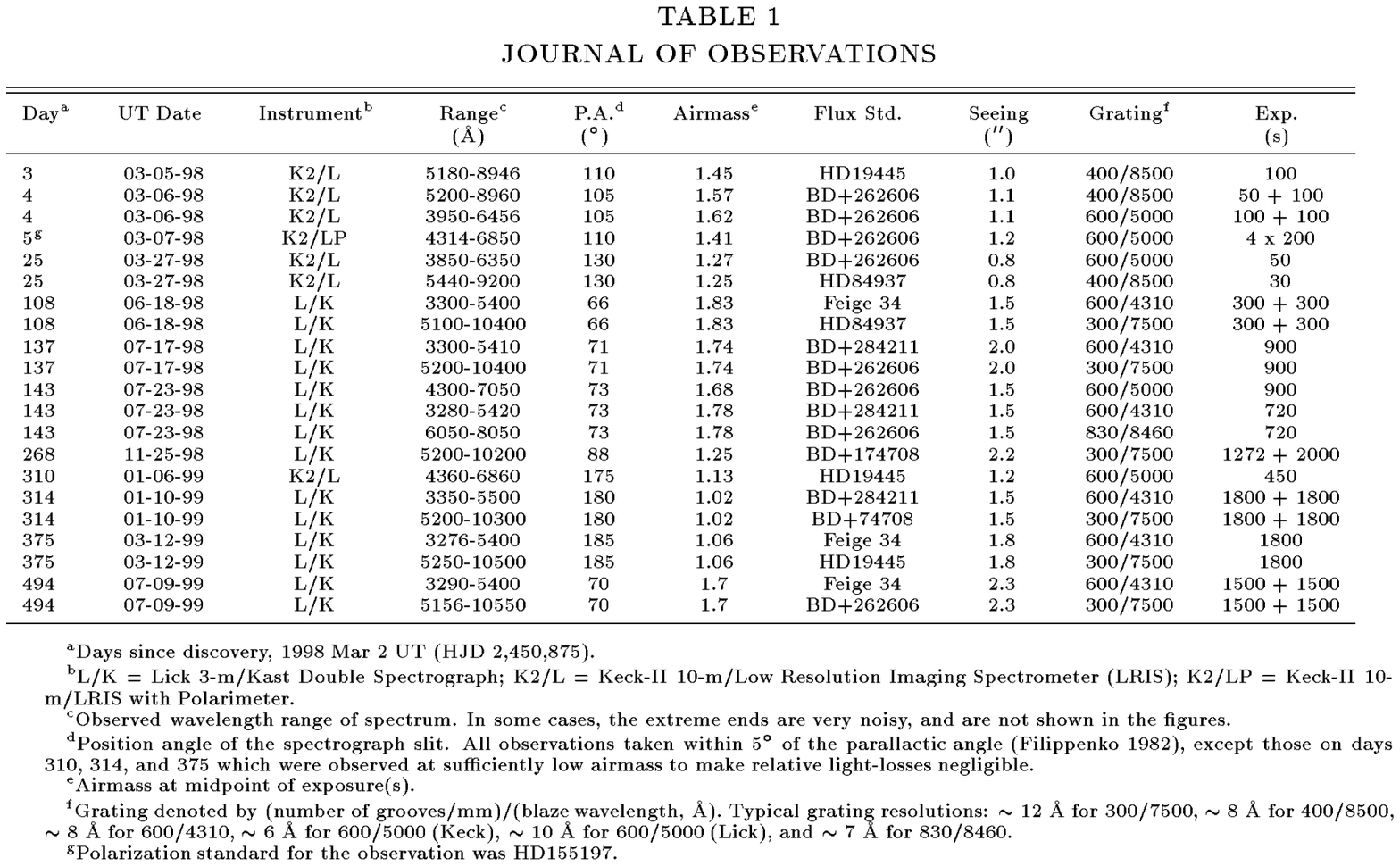}
		}
\end{center}
\end{figure}

\newpage

\begin{figure}[ht!]
\begin{center}

 \scalebox{.9}{
	\includegraphics{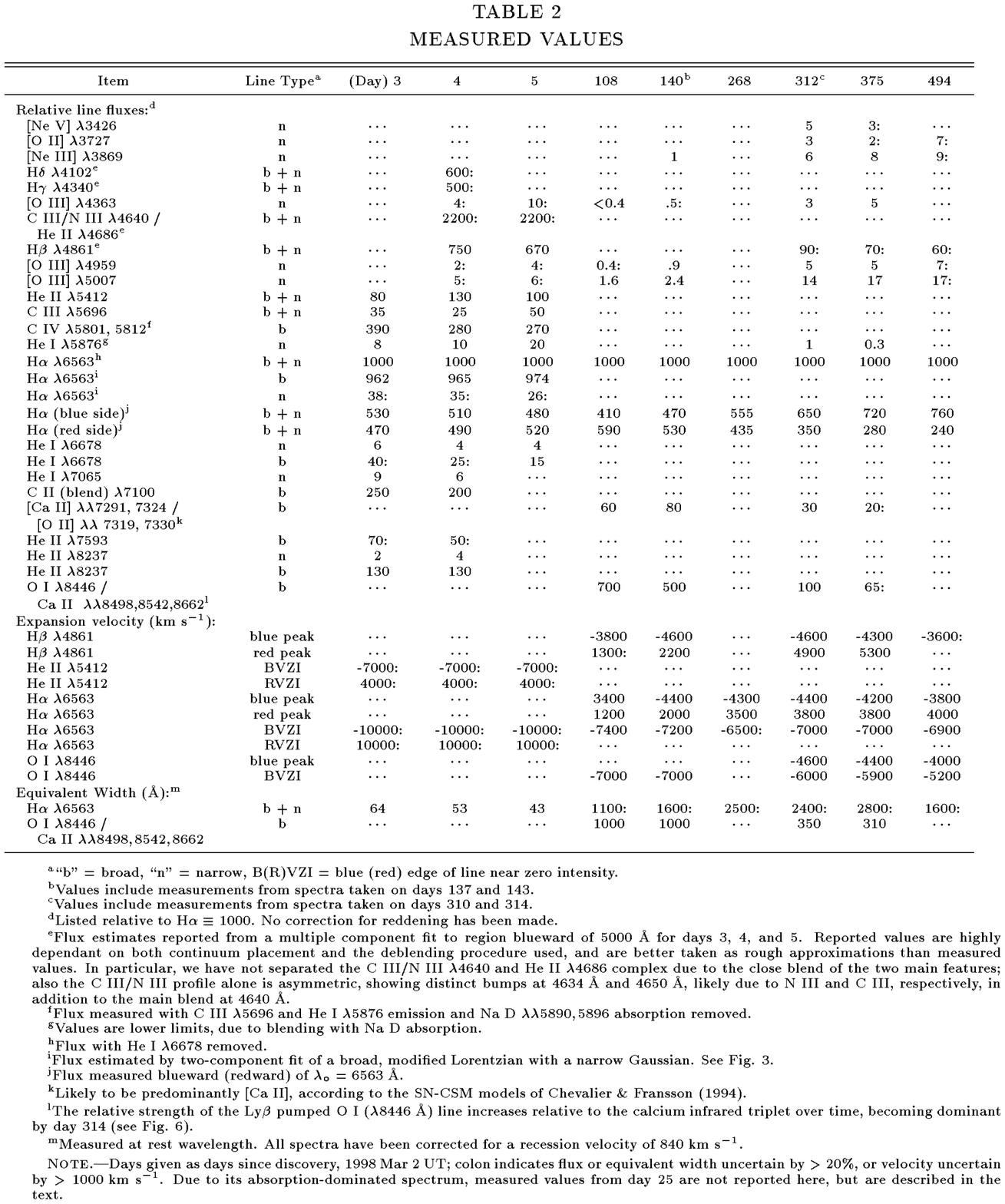}
		}
\end{center}
\end{figure}

\newpage

\begin{figure}[ht!]
\begin{center}

 \scalebox{1.0}{
	\includegraphics{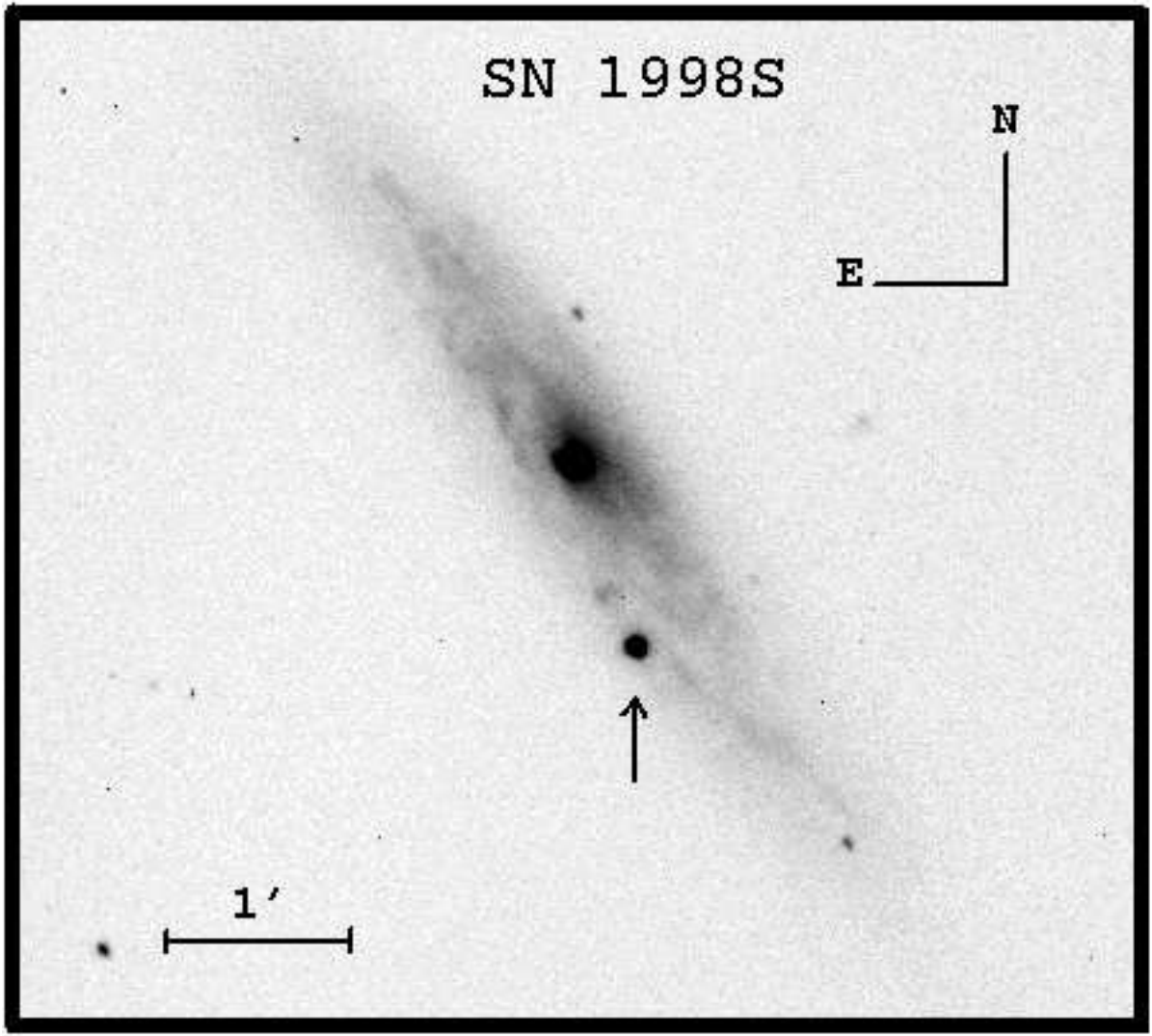}
		}
\end{center}
\caption{$V$-band image of NGC 3877 taken on 1998 March 6 with the Katzman
Automatic Imaging Telescope (Treffers et al. 1997).  SN 1998S (arrow) is about
16$^{\prime\prime}$ west and 46$^{\prime\prime}$ south of the nucleus (Li et
al. 1998). }
\end{figure}

\newpage

\begin{figure}[ht!]
\begin{center}
 \rotatebox{90}{
 \scalebox{0.7}{
	\includegraphics{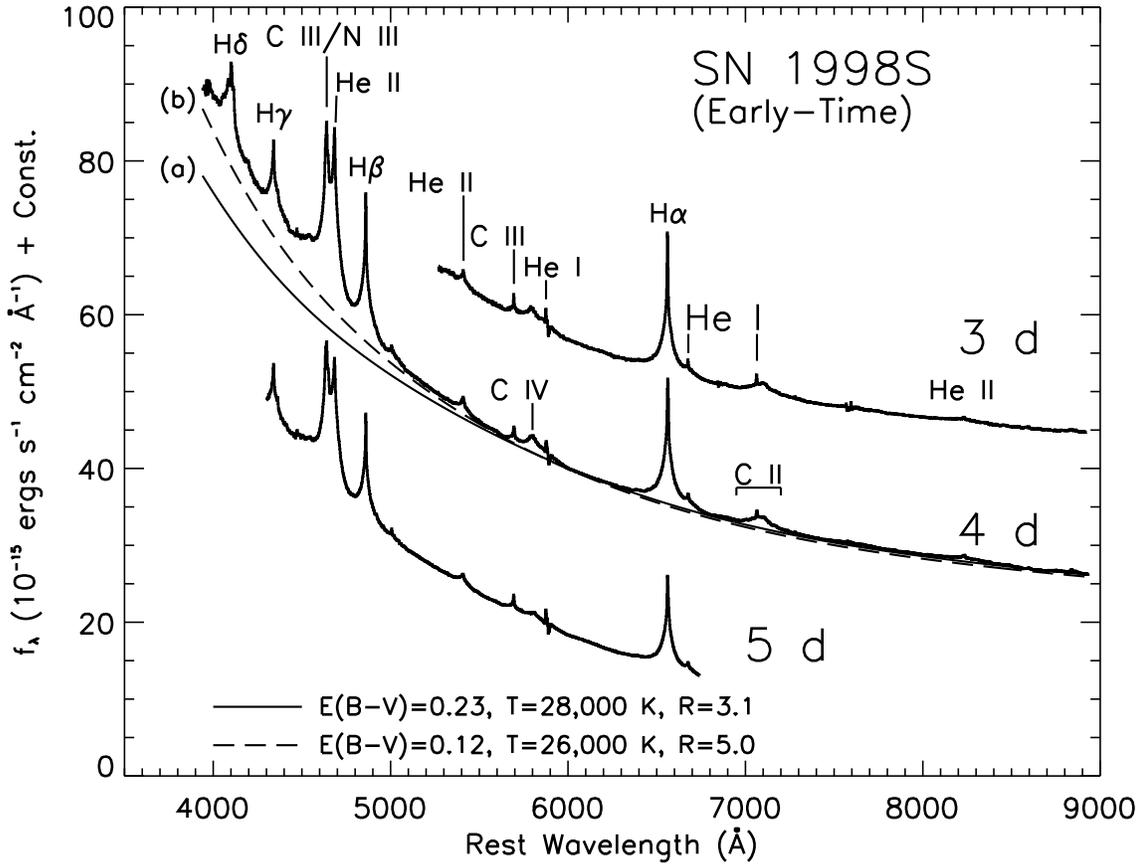}
		}
	        }
\end{center}
\caption{The early-time spectral development of SN 1998S.  Dates are
relative to date of discovery, 1998 March 2.  Constants of 40 and 20 have been
added to the day 3 and day 4 spectra, respectively.  Note that in this and all
figures the spectra have been corrected for a recession velocity of 840 km
s$^{-1}$, determined from the average of the narrow \ion{He}{1} lines.  Fitting
the day 4 continuum with a reddened blackbody [$E{\rm (}B-V{\rm )} = 0.23$ with
$R_V=3.1$] and allowing $T$ to vary freely produces the best fit labeled ({\it
a}); the excess flux beyond a thermal continuum apparent at blue wavelengths
(see also Fig. 4) remains for any reddening or temperature assumed.  Either a
non-standard extinction law [$R_V\gtrsim 4.0$; line ({\it b})] or an intrinsic
blue excess can explain this behavior. See text for details.}

\end{figure}

\newpage

\begin{figure}[ht!]
\begin{center}

 \scalebox{0.7}{
	\includegraphics{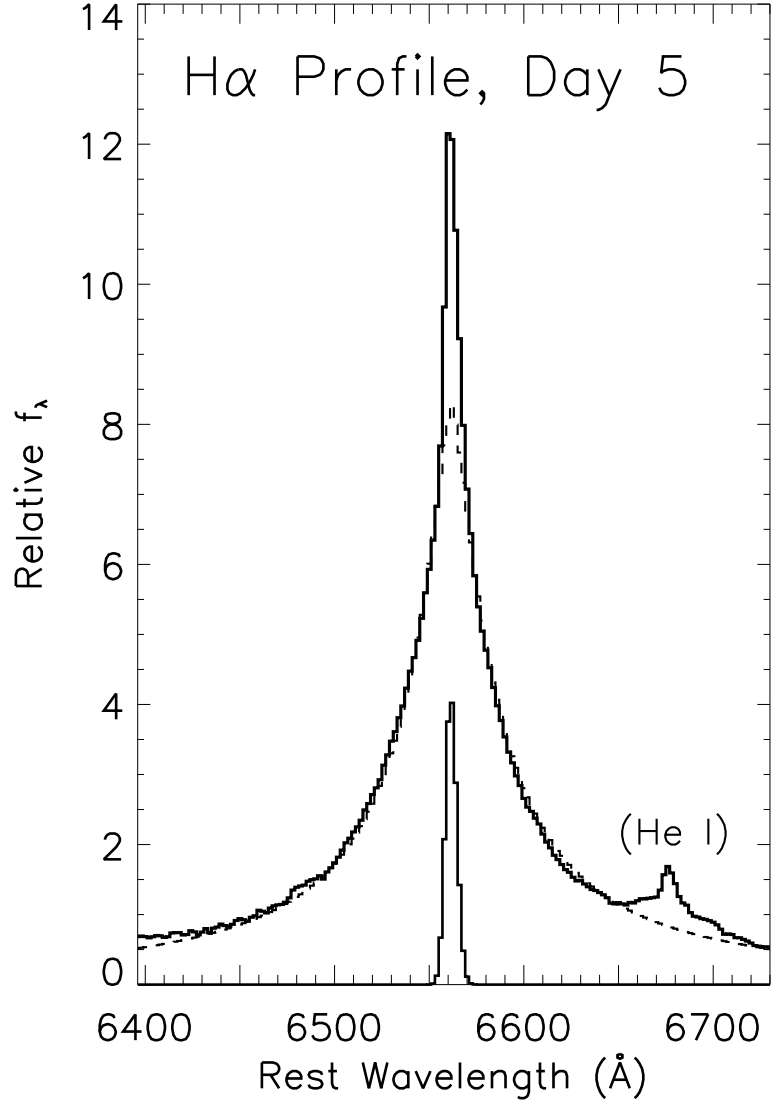}
		}
\end{center}
\caption{A two-component fit to the H$\alpha$ profile ({\it thick solid
line}) on 1998 March 7, five days after discovery.  The best fits were achieved
with the combination of a broad, modified ($f_\lambda\propto\ \mid \lambda -
\lambda_\circ \mid ^\alpha;\ \alpha=-1.4$) Lorentzian ({\it dashed line}) and
an unresolved (FWHM $\lesssim 300\ {\rm km\ s}^{-1}$) Gaussian ({\it thin solid
line}).  Fits using only a single modified Lorentzian component were less
successful; spectropolarimetry and later-time spectra provide additional
evidence for the narrow component.}
\end{figure}

\newpage


\begin{figure}[ht!]
\begin{center}

 \scalebox{0.75}{
	\includegraphics{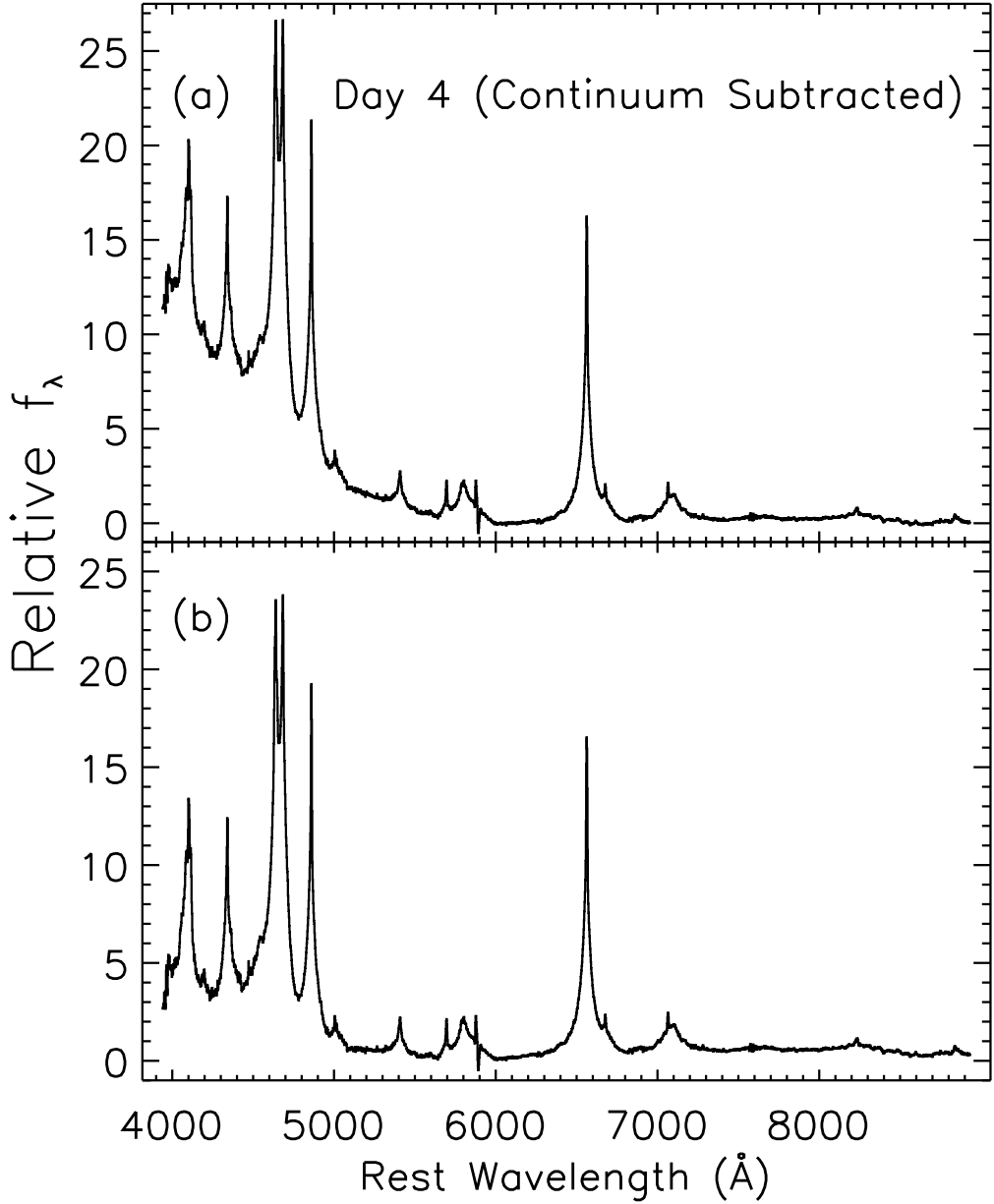}
		}
\end{center}
\caption{SN 1998S on 1998 March 6, four days after discovery, with the
reddened, thermal continua labeled in Fig. 2 subtracted. While the excess flux
seen at blue wavelengths in $(a)$ (day 4 spectrum with the model continuum
labeled $(a)$ in Fig. 2 subtracted) could possibly be explained by an unusual
extinction law (demonstrated in $(b)$, where the day 4 spectrum with the model
continuum labeled $(b)$ in Fig. 2 has been subtracted), it is more likely
intrinsic to the SN itself (see \S 4.2).}
\end{figure}

\newpage

\begin{figure}[ht!]
\begin{center}
 \rotatebox{90}{
 \scalebox{0.7}{
	\includegraphics{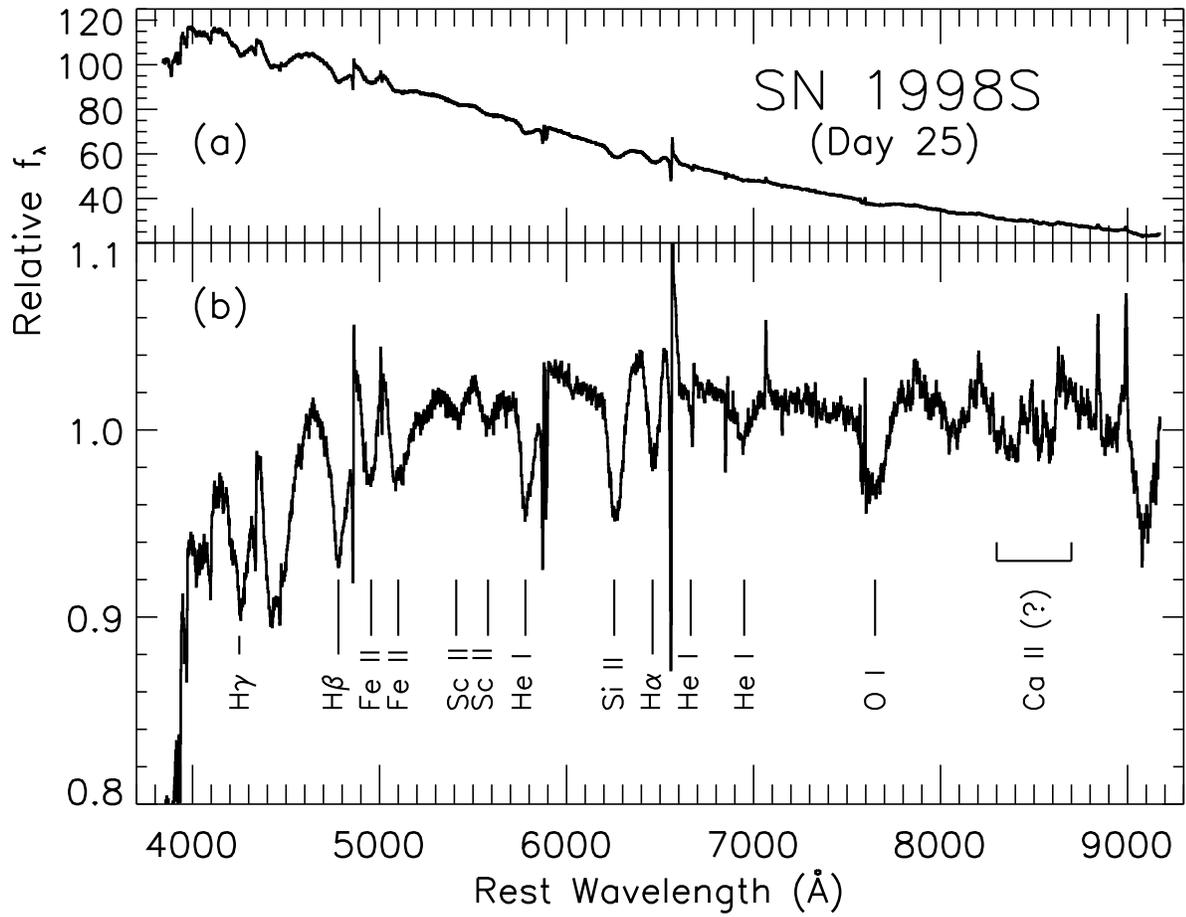}
		}
		}
\end{center}
\caption{{\it (a)} Spectrum of SN 1998S on 1998 March 27, 25 days after
discovery. {\it (b)} The spectrum normalized by division with a T=10,000 K
blackbody ($E(B-V)=0.1,\ R_V = 3.1$).  Probable absorption line identifications
are indicated.}
\end{figure}

\newpage

\begin{figure}[ht!]
 \rotatebox{90}{
 \scalebox{0.7}{
	\includegraphics{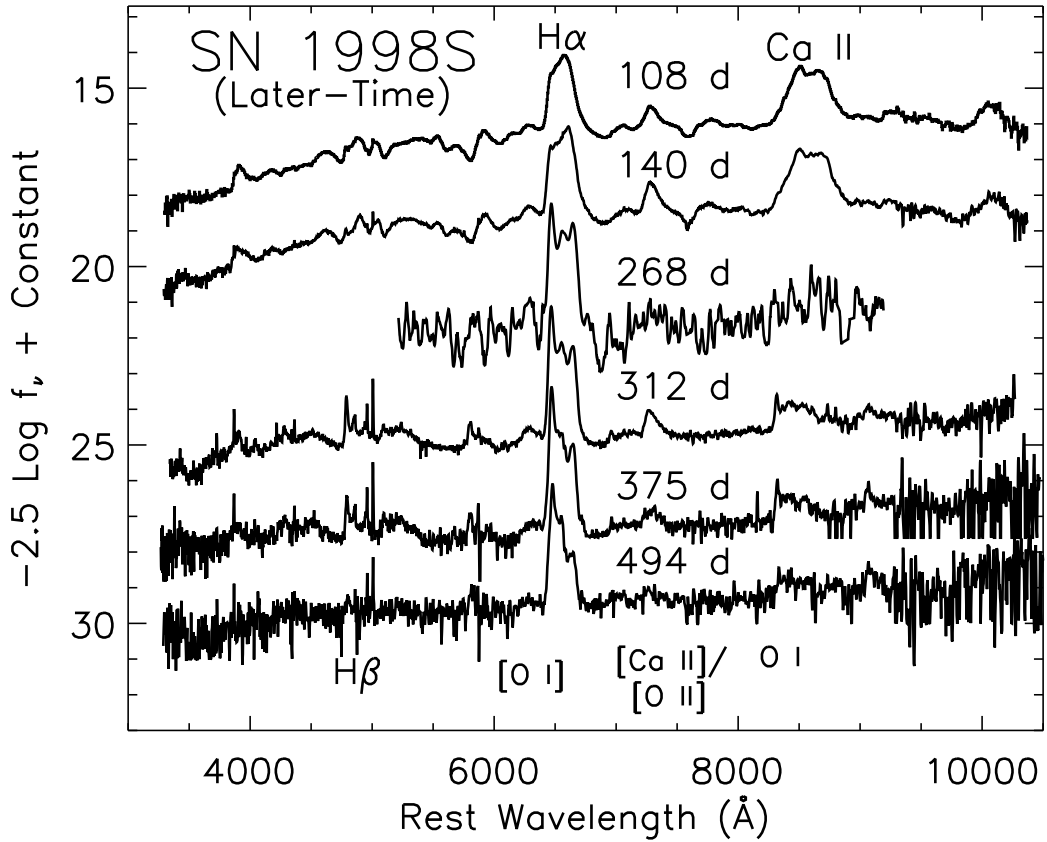}
		}
		}
\caption{The later-time spectral development of SN 1998S.  Epochs (days)
given relative to the discovery date, 1998 March 2.  The spectrum at 140 days
is the combination of spectra taken 1998 July 17 \& 23, and the spectrum at 312
days is the combination of spectra taken 1999 January 6 \& 10.  Due to its poor
signal-to-noise ratio, the spectrum from day 268 has been smoothed with a
second degree Savitzky-Golay smoothing filter (Press et al. 1992) thirty pixels
wide.}
\end{figure}

\begin{figure}[ht!]
\begin{center}
 \rotatebox{0}{
 \scalebox{0.8}{
	\includegraphics{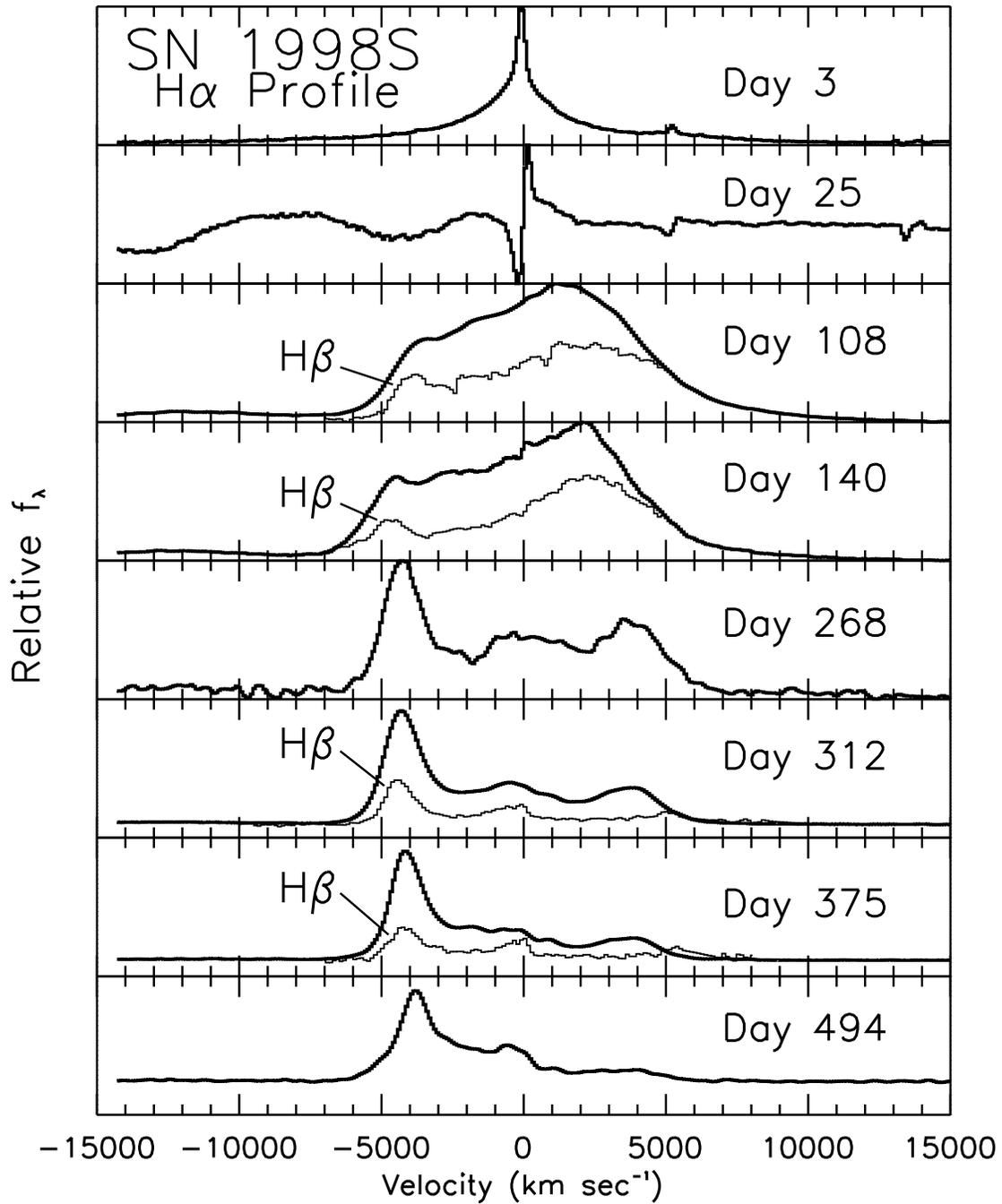}
		}
		}
\end{center}
\caption{Evolution of the H$\alpha$ line.  Epochs (days) given relative to
the discovery date, 1998 March 2.  The H$\beta$ profile, scaled by arbitrary
amounts for comparison of profile shape, is overplotted in four of the
later-time spectra.  }
\end{figure}

\newpage

\begin{figure}[ht!]
\begin{center}
 \rotatebox{0}{
 \scalebox{0.7}{
	\includegraphics{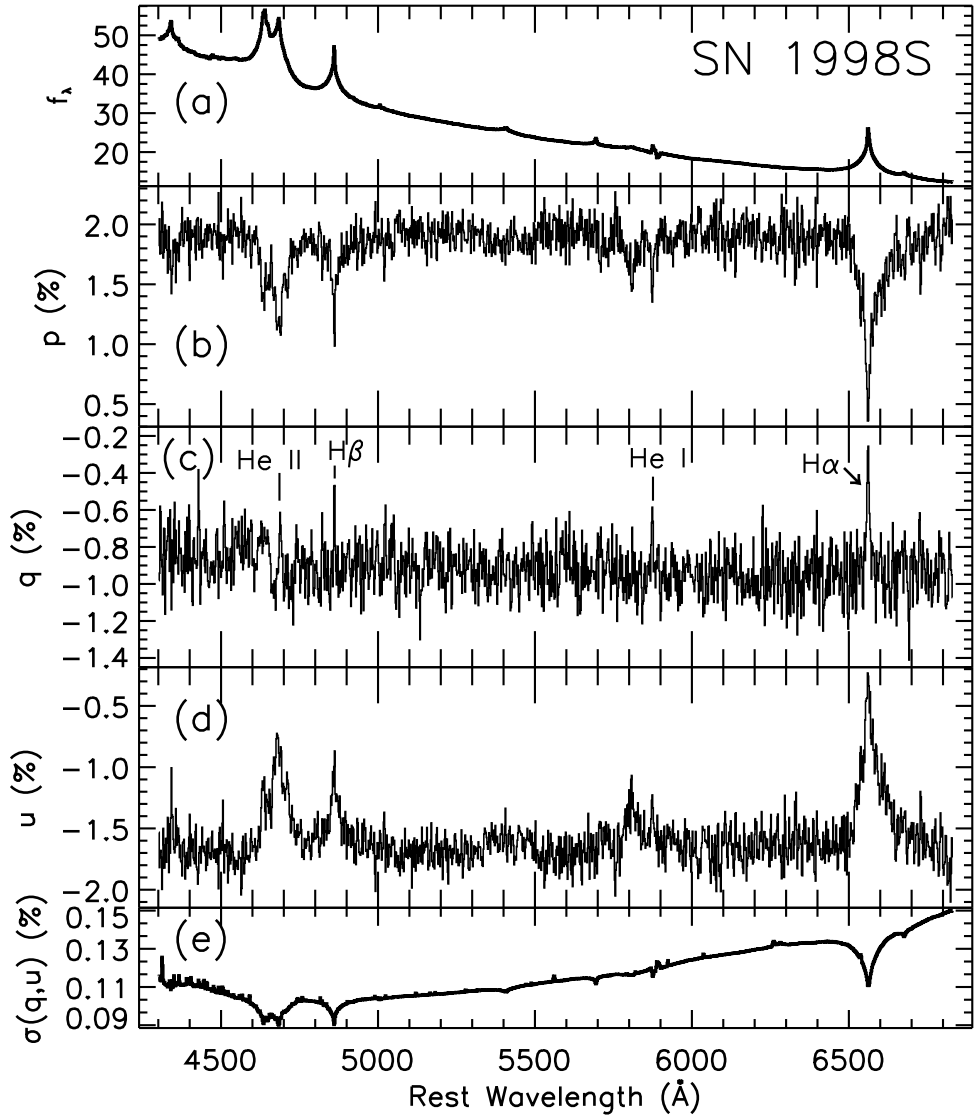}
		}
		}
\end{center}
\caption{Polarization data for SN 1998S, obtained 1998 March 7, 5 days after
discovery.  {\it (a)} Total flux, in units of $10^{-15}$ ergs s$^{-1}$ cm$^{-2}$
\AA$^{-1}$.  {\it (b)} Observed degree of polarization.  {\it (c, d)} The
normalized $q$ and $u$ Stokes parameters, with prominent narrow-line features
identified. {\it (e)} Average of the (nearly identical) $1\sigma$ uncertainties
(statistical) in the Stokes $q$ and $u$ parameters.  Note that the polarization
shown in this and all plots is actually the ``rotated Stokes parameter''
($RSP$; see Tran 1995), though the high polarization of this object renders the
difference between $p$ and $RSP$ nearly negligible. }
\end{figure}

\newpage

\begin{figure}[ht!]
\begin{center}
 \rotatebox{0}{
 \scalebox{0.85}{
	\includegraphics{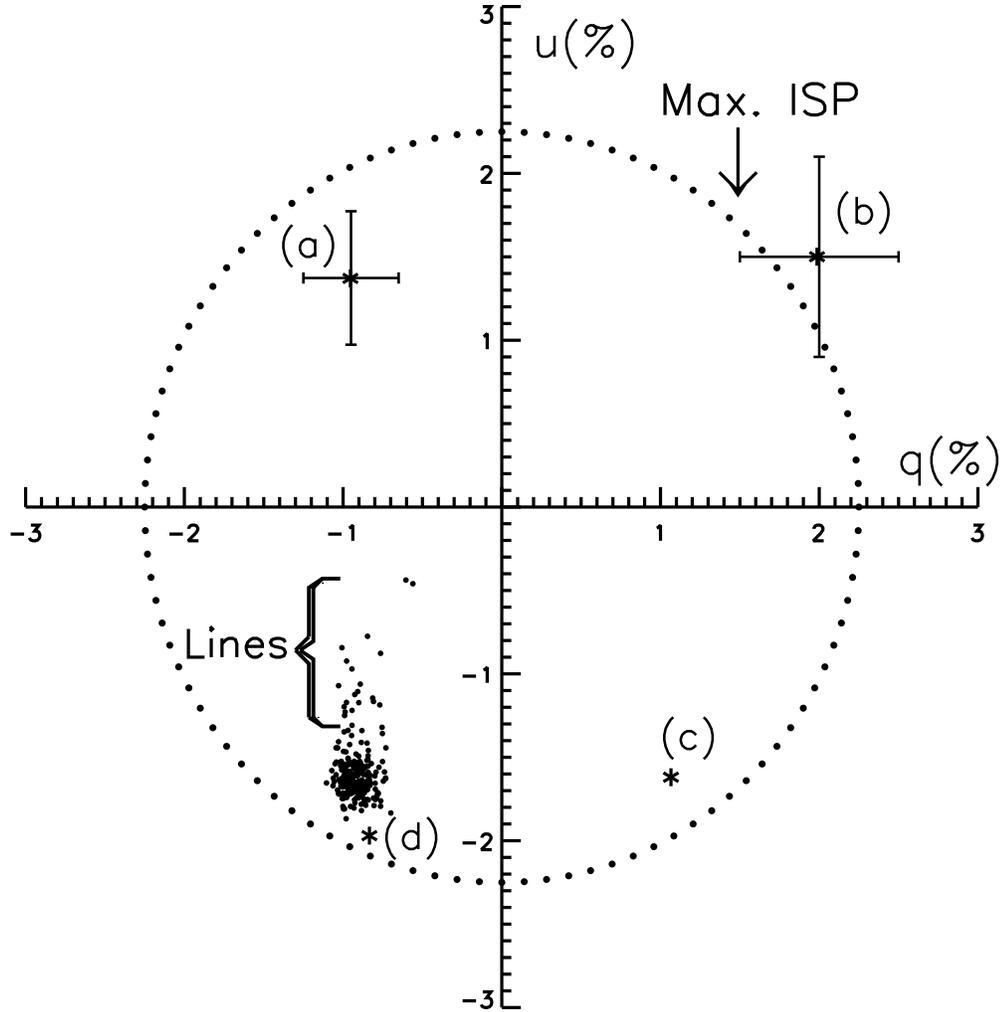}
		}
		}
\end{center}
\caption{Polarization data in the $q$-$u$ plane for SN 1998S on 1998 March
7.  Each point represents a bin 10 \AA\ wide.  Also shown is the maximum ISP
inferred from the estimated reddening limit (\S3.2).  The four labeled
asterisks represent ISP values which, when removed from the total polarization,
produce the ``intrinsic'' SN polarizations shown in Fig. 10.  Point ({\it a})
represents the ISP resulting from the model of a polarized continuum surrounded
by an unpolarized broad-line region (\S3.3.3); a similar model, but with
unpolarized narrow lines (\S3.3.4), results in the ISP indicated by point ({\it
b}).  The error bars on points ({\it a}) and ({\it b}) reflect the range of ISP
values allowed under the two models.  Points ({\it c}) and ({\it d}) are
allowable ISP values if neither the broad nor the narrow lines are unpolarized
(\S3.3.5).  }
\end{figure}

\newpage

\begin{figure}[ht!]
\begin{center}
 \rotatebox{0}{
 \scalebox{0.8}{
	\includegraphics{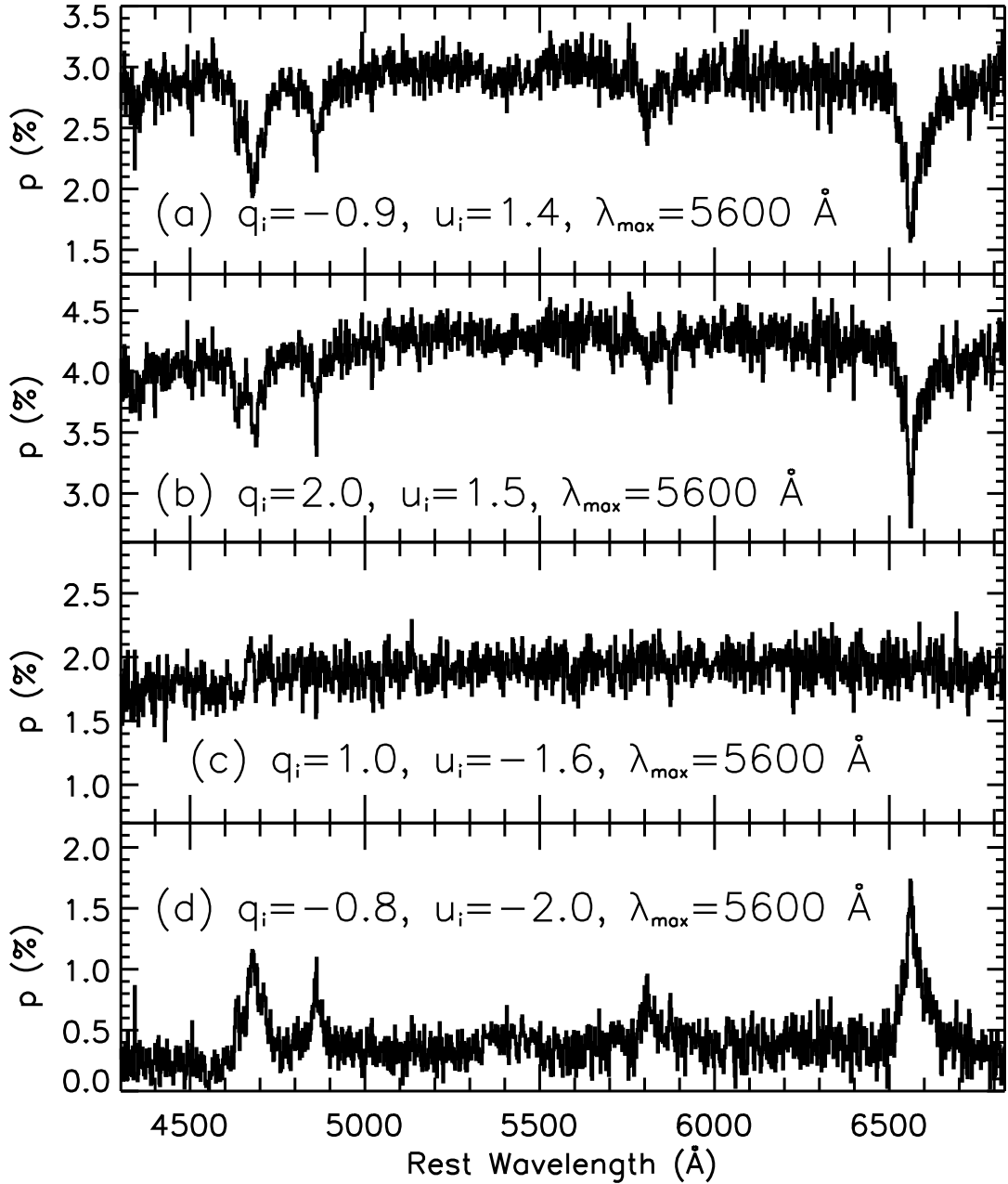}
		}
		}
\end{center}
\caption{The effects of uncertain ISP removal: The ``intrinsic'' SN
polarization resulting from the four ISP values labeled in Fig. 9.  Based on
the model of a polarized continuum with depolarizing broad-line features
described in the text (\S3.3.3), {\it(a)} is considered the most likely
representation of the polarization intrinsic to SN 1998S. }
\end{figure}

\newpage

\begin{figure}[ht!]
\begin{center}
 \rotatebox{0}{
 \scalebox{0.8}{
	\includegraphics{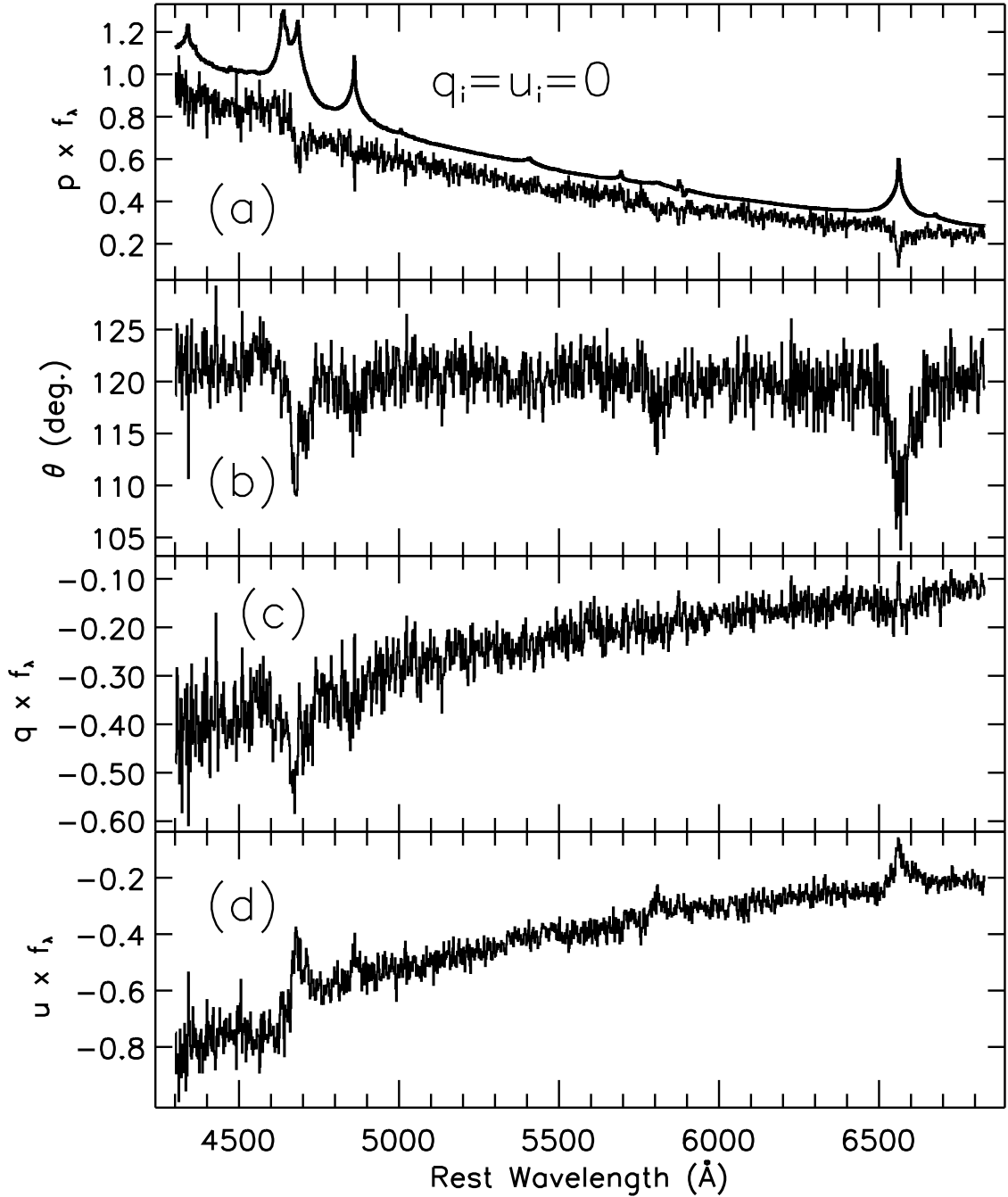}
		}
		}
\end{center}
\caption{ {\it (a)} The Stokes flux (equal to $p \times f_{\lambda}$), {\it
(b)} polarization PA $(\theta)$, and {\it (c, d)} Stokes parameter fluxes
($q\times f_\lambda {\rm\ or\ } u \times f_{\lambda}$) for SN 1998S on 1998
March 7 assuming zero ISP.  The total flux spectrum, scaled by 0.023, is
shown for comparison of features in {\it (a)}. }
\end{figure}

\newpage

\begin{figure}[ht!]
\begin{center}
 \rotatebox{0}{
 \scalebox{0.65}{
	\includegraphics{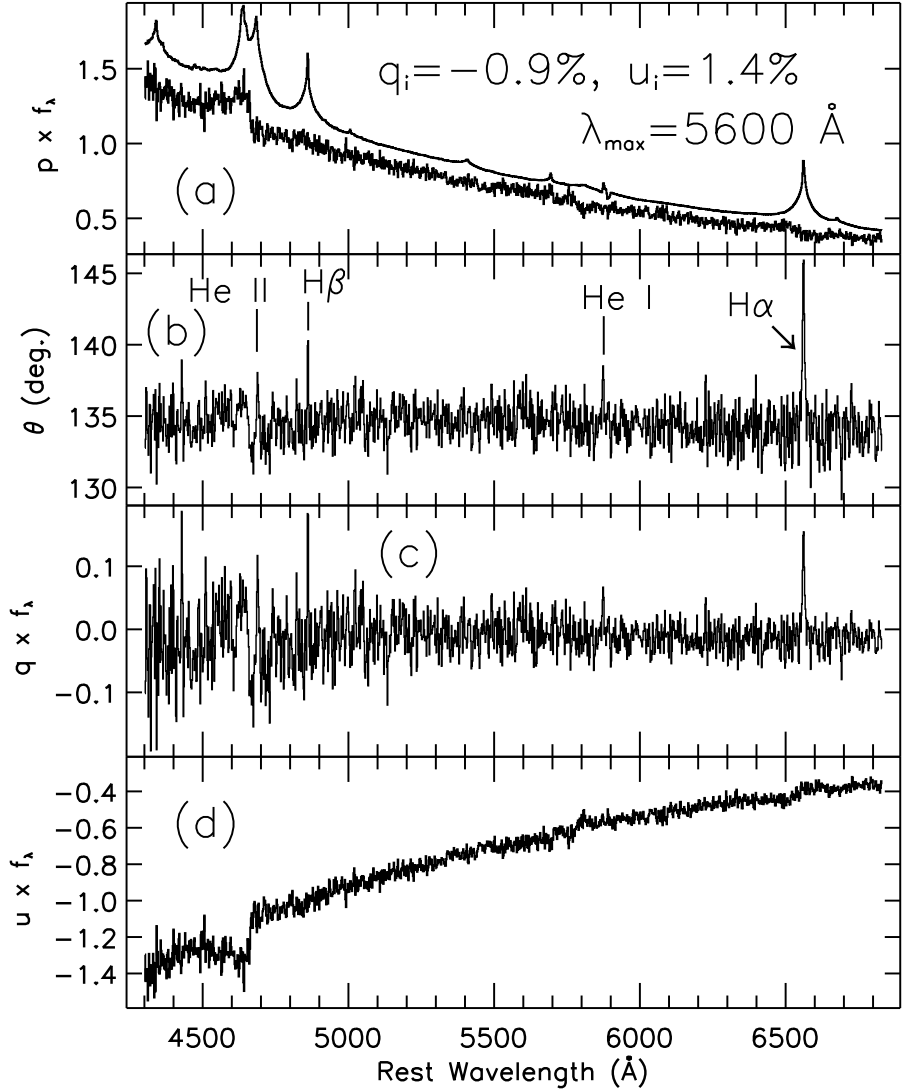}
		}
		}
\end{center}
\caption{ {\it (a)} The Stokes flux, {\it (b)} polarization PA, and {\it
(c, d)} Stokes parameter fluxes for SN 1998S on 1998 March 7 assuming the ISP
derived from the model of a polarized continuum and unpolarized broad-line
features described in the text (\S3.3.3).  The total flux spectrum, scaled by
0.034, is shown for comparison of features in {\it (a)}.  Note that except for
the \ion{C}{3}/\ion{N}{3} $\lambda4640$ line, the broad polarization features
seen prior to ISP removal (Fig. 11) have essentially disappeared; the increase
in polarized flux evident in the \ion{C}{3}/\ion{N}{3} line suggests that it
originates from a slightly different physical region than the other lines.  The
prominent narrow-line features seen in both the PA (labeled) and the $q$ Stokes
flux are interpreted in this model as resulting from additional scattering by
asymmetrically distributed CSM. }
\end{figure}

\newpage

\begin{figure}[ht!]
\begin{center}
 \rotatebox{0}{
 \scalebox{0.75}{
	\includegraphics{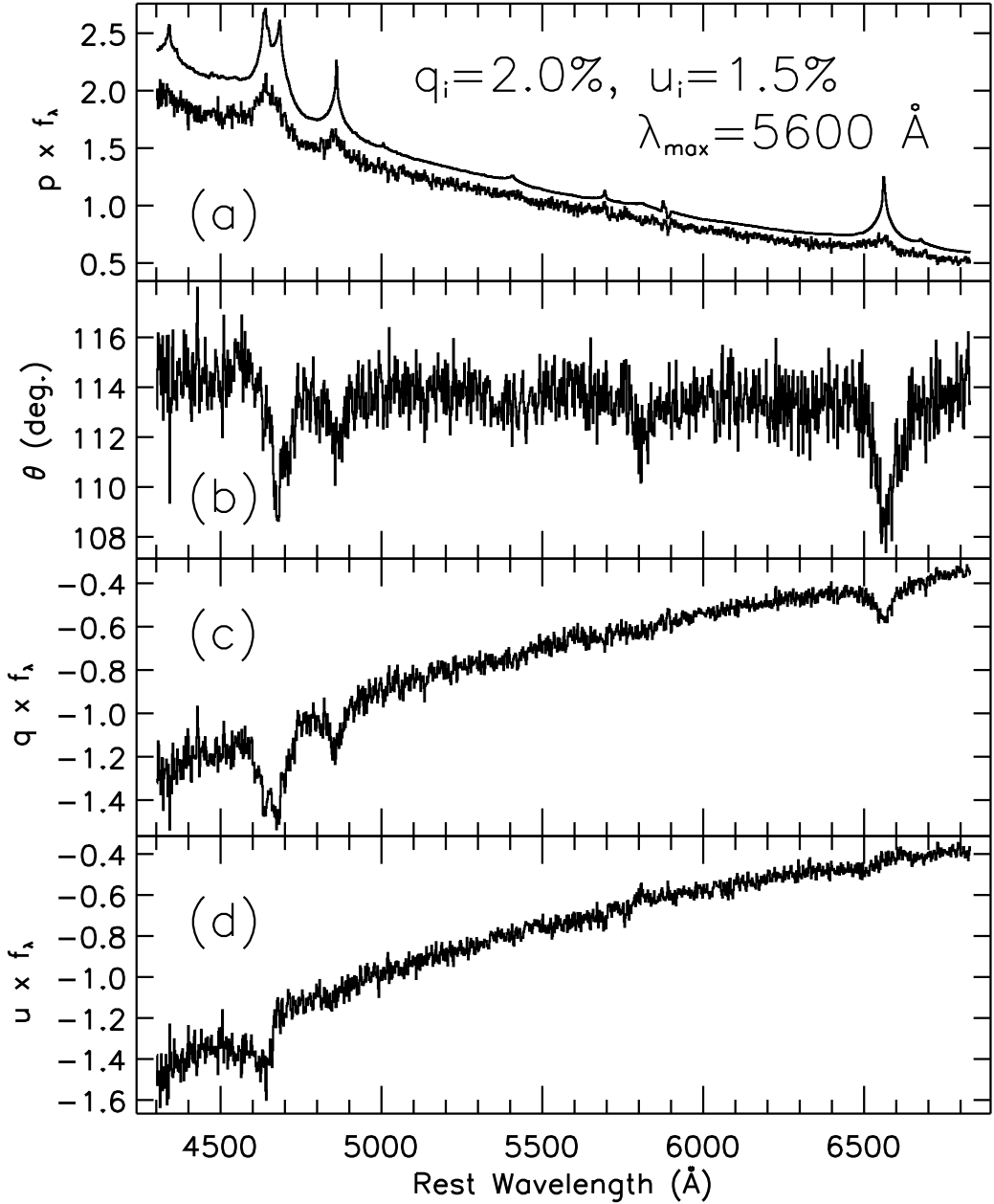}
		}
		}
\end{center}
\caption{{\it (a)} The Stokes flux, {\it (b)} polarization PA, and {\it (c,
d)} Stokes parameter fluxes for SN 1998S on 1998 March 7 assuming the ISP
derived from the model of a polarized continuum and unpolarized narrow-line
features described in the text (\S3.3.4).  The total flux spectrum, scaled by
0.048, is shown for comparison in {\it (a)}.  In this model, the polarization
change and PA rotation seen across the broad-line features result from an
axially symmetric scattering environment with a geometry different from the
continuum.   }
\end{figure}

\newpage

\begin{figure}[ht!]
\begin{center}
 \rotatebox{270}{
 \scalebox{0.75}{
	\includegraphics{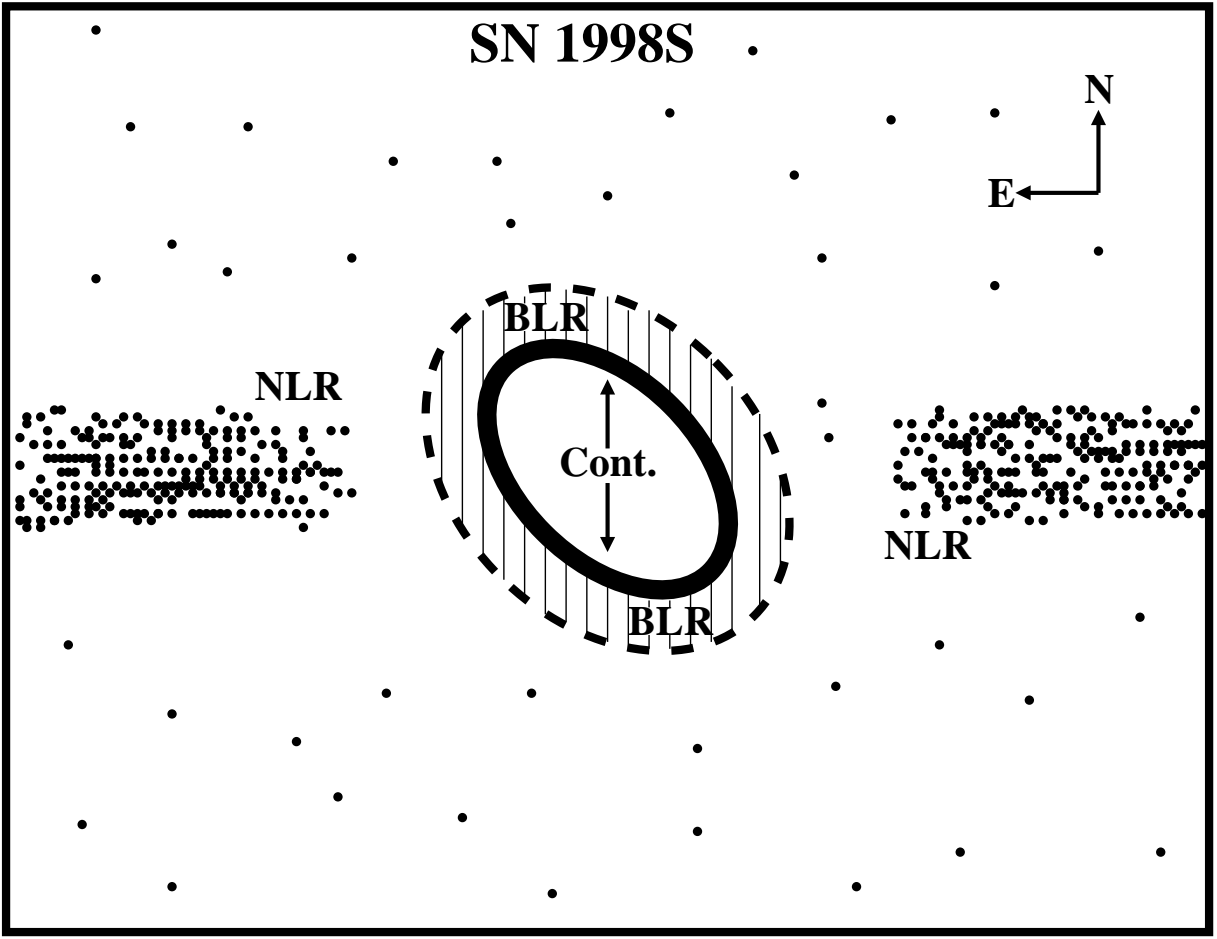}
		}
		}
\end{center}
\caption{A possible geometry of SN 1998S and its CSM, five days after
discovery, showing schematic (not to scale) representations of the regions
responsible for the continuum (Cont.), broad lines (BLR), and narrow lines
(NLR).  This morphology and its orientation in the plane of the sky were
derived by assuming unpolarized broad lines (or slightly polarized, but with
the same geometry as the continuum-forming region) surrounding a polarized
continuum (\S 3.3.3).  We have represented the NLR as being concentrated in a
disk (arbitrarily shown edge-on), a geometry suggested by later-time total flux
spectra.  The representation of the continuum source as a shell of material
results from theoretical considerations described in the text (\S 4.1).}
\end{figure}

\end{document}